\documentclass[fleqn,usenatbib]{mnras}

\usepackage[T1]{fontenc}
\usepackage{ae,aecompl}

\usepackage{graphicx}	% Including figure files
\usepackage{epstopdf}
\usepackage{amsmath}	% Advanced maths commands
\usepackage{amssymb}	% Extra maths symbols
\usepackage{multirow,array}
\usepackage{soul}
\usepackage[usenames, dvipsnames]{color}  % Coloured text etc.
\usepackage[usenames, dvipsnames]{xcolor}  % Coloured text etc.
\usepackage{comment}
\usepackage{caption}
\usepackage{subcaption}
\usepackage{float}
\usepackage{newtxtext,newtxmath}

\usepackage{soul}
\title[DeepDISC: Detectron2]{Detection, Instance Segmentation, and Classification for Astronomical Surveys with Deep Learning (DeepDISC): Detectron2 Implementation and Demonstration with Hyper Suprime-Cam Data}

\author[Merz et al.]{
Grant Merz,$^{1}$\thanks{E-mail: gmerz3@illinois.edu}
Yichen Liu,$^{1}$
Colin J. Burke,$^{1}$
Patrick D. Aleo,$^{1}$
Xin Liu,$^{1,2,3}$
Matias Carrasco Kind,$^{1,2}$
\newauthor
Volodymyr Kindratenko,$^{2,3,4,5}$
Yufeng Liu$^{6}$
\\
$^{1}$Department of Astronomy, University of Illinois at Urbana-Champaign, 1002 West Green Street, Urbana, IL 61801, USA \\
$^{2}$National Center for Supercomputing Applications, University of Illinois at Urbana-Champaign, 1205 West Clark Street, Urbana, IL 61801, USA \\
$^{3}$Center for Artificial Intelligence Innovation, University of Illinois at Urbana-Champaign, 1205 West Clark Street, Urbana, IL 61801, USA \\
$^{4}$Department of Computer Science, University of Illinois at Urbana-Champaign, 201 North Goodwin Avenue, Urbana, IL 61801, USA \\
$^{5}$Department of Electrical and Computer Engineering, University of Illinois at Urbana-Champaign, 306 North Wright Street, Urbana, IL 61801, USA \\
$^{6}$Department of Physics, University of Illinois at Urbana-Champaign, 1110 West Green Street, Urbana, IL 61801, USA \\
}

\date{Accepted XXX. Received YYY; in original form ZZZ}

\pubyear{2023}

\begin{document}
\label{firstpage}
\pagerange{\pageref{firstpage}--\pageref{lastpage}}
\maketitle

\begin{abstract}
The next generation of wide-field deep astronomical surveys will deliver unprecedented amounts of images through the 2020s and beyond. As both the sensitivity and depth of observations increase, more blended sources will be detected. This reality can lead to measurement biases that contaminate key astronomical inferences.  We implement new deep learning models available through Facebook AI Research's Detectron2 repository to perform the simultaneous tasks of object identification, deblending, and classification on large multi-band coadds from the Hyper Suprime-Cam (HSC). We use existing detection/deblending codes and classification methods to train a suite of deep neural networks, including state-of-the-art transformers. Once trained, we find that transformers outperform traditional convolutional neural networks and are more robust to different contrast scalings. Transformers are able to detect and deblend objects closely matching the ground truth, achieving a median bounding box Intersection over Union of 0.99. Using high quality class labels from the Hubble Space Telescope, we find that when classifying objects as either stars or galaxies, the best-performing networks can classify galaxies with near 100\% completeness and purity across the whole test sample and classify stars above 60\% completeness and 80\% purity out to HSC i-band magnitudes of 25 mag. This framework can be extended to other upcoming deep surveys such as the Legacy Survey of Space and Time and those with the Roman Space Telescope to enable fast source detection and measurement.  Our code, \textsc{DeepDISC}, is publicly available at \url{https://github.com/grantmerz/deepdisc}.
\end{abstract}

\begin{keywords}
techniques: image processing -- methods: data analysis -- galaxies: general -- stars: general
\end{keywords}

\section{Introduction}

\label{sec:intro}

The rise of machine learning/artificial intelligence has allowed for rapid advancement in many image analysis tasks to the benefit of researchers who wish to work with large sets of imaging data. This active field of study, known as computer vision, has led to developments in many disciplines including medical imaging \citep{Zhou20}, urban planning \citep{IBRAHIM20}, autonomous systems \citep{Pavel22} and more. 

 Tasks such as image compression, inpainting, object classification and detection, and many others have been extensively studied. Astronomy is no exception, and many methods that utilize deep learning have been applied to simulations and real survey data for tasks such as object detection, star/galaxy classification, photometric redshift estimation, image generation, deblending and more (see \citealt{Dawes23} for a comprehensive review).  Machine learning methods are already becoming instrumental in handling the large volume of data processed every day in survey pipelines \citep[e.g.,][]{bosch_hyper_2018, Russeil22, Tachibana18, Malanchev21, Mahabal19}

The next generation of astronomical surveys such as the upcoming Legacy Survey of Space and Time \citep[LSST;][]{Ivezic2019} at the Vera C. Rubin Observatory, the Wide-Field Imaging Survey at the Nancy Grace Roman Space Telescope \citep[\textit{Roman};][]{Spergel13}, and  \textit{Euclid} \citep{Amiaux12} will produce unprecedented amounts of imaging data throughout the 2020s and beyond.  LSST will provide incredibly deep ground-based observations of the sky, revealing a map of the universe including objects as faint as $\sim$25-27 mag at a 5$\sigma$ detection for 10 year observing runs.  Ground-based surveys such as the Hyper Suprime-Cam Subaru Strategic Program \citep[HSC SSP;][]{aihara_hyper_2018} and the Dark Energy Survey \citep[DES;][]{DES16} have already mapped large swaths of the sky and produced catalogs of tens of millions of objects, with HSC depths being comparable to LSST. The astronomical research community is now in an era that demands robust and efficient techniques to detect and analyze sources in images. 
 
Current surveys such as HSC already report large fractions of blended (overlapping) objects. For instance, 58\% of objects in the the shallowest field (Wide) of the HSC survey are blended, i.e., detected in a region of sky above the 5$\sigma$ threshold (26.2 mag) containing multiple significant peaks in surface brightness.  As depths increase, line-of-sight projections and physical mergers cause the overall number of blends to increase.  This fraction rises to 66\% for the Deep and 74\% for the UltraDeep layers, which are comparable to LSST depths \citep{bosch_hyper_2018}. If blends are not identified, they will bias results from pipelines that assume object isolation. For example, \cite{Boucaud19} show that the traditional detection/deblending methods can lead to a photometric error of >0.75 mag for $\sim$12\% of their sample of artificially blended galaxies from the Cosmic Assembly Near-infrared Deep Extragalactic
Legacy survey \citep[CANDELS][]{CANDELS1, CANDELS2}. Unrecognized blends can cause an increase in the noise of galaxy shear measurements by $\sim$14\% for deep observations \citep{Dawson16}.  Deblending, or source separation, has been recognized as a high priority in survey science, especially as LSST begins preparations for first light.

Despite rigorous efforts to deblend objects, the problem of deblending remains, and in some sense will always remain in astronomical studies.  Deblending involves separating a mixture of signals in order to independently measure properties of each individual object. This an imaging problem analogous to the ``cocktail party problem'', in which an attempt is made to isolate individual voices from a mixture of conversations.  However, since it is impossible to trace a photon back to an individual source, astronomical deblending is characterized as an under-constrained inverse problem. Deblending methods must rely on assumptions about source properties and models of signal mixing \citep{melchior_challenge_2021}.

A first step in deblending is object detection.  Many codes have been developed for source detection and classification, including $\textsc{FOCAS}$ \citep{Jarvis81}, $\textsc{NEXT}$ \citep{Andreon00} and $\textsc{SExtractor}$ \citep{SExtractor}.  $\textsc{SExtractor}$ is widely used in survey pipelines including HSC \citep{bosch_hyper_2018} and DES \citep{Morganson2018}, but can be sensitive to configuration parameters.  While $\textsc{SExtractor}$ also deblends by segmenting, or identifying pixels belonging to unique sources, modern deblenders have been developed such as $\textsc{Morpheus}$ \citep{hausen_morpheus_2020} and  $\textsc{Scarlet}$, \citep{melchior_scarlet_2018} with the latter implemented in HSC and LSST pipelines.  With hopes for real-time object detection and deblending algorithms in surveys such as LSST, machine learning applications to crowded fields offer a promising avenue.  The use of deep neural networks, or deep learning has seen particular success in image processing. In addition to efficiency and flexibility, neural networks may be able to overcome limitations of traditional peak-finding algorithms due to their fundamentally different detection mechanism.

There is a growing body of deep learning deblending methods in astronomy. \cite{reiman_deblending_2019} use a Generative Adversarial Network (GAN) to deblend small cutouts of Sloan Digital Sky Survey \citep[SDSS][]{SDSS} galaxies from Galaxy Zoo \citep{Lintott11}. \cite{arcelin_deblending_2021} use a variational autoencoder to deblend small cutouts of simulated LSST galaxies. \cite{Hemmati22} use GANs to deblend images with HSC resolution and recover Hubble Space Telescope resolution.  On larger scales, \cite{bretonniere_probabilistic_2021} use a probabilistic U-net model to deblend large simulated scenes of galaxies.

In addition to blending, another pressing issue with increased depth is the presence of many unresolved galaxies in the deep samples of smaller and fainter objects.  This will prove difficult for star-galaxy classification schemes that rely on morphological features to distinguish between a point source star or a point source galaxy, although machine learning methods have been employed to combat this problem  \citep{Tachibana18, Miller21}. \cite{Muyskens22} use a Gaussian process classifier to perform star/galaxy classification on HSC images. This is an important area of study, as misclassifications can introduce biases in studies that require careful measurement of galaxy properties. For instance, it has been shown that stellar contamination can be a significant source of bias in galaxy clustering measurements \citep{Ross11}.  Precise constraints of cosmological models require a correction of this systematic bias in measurements of clustering at high photometric redshifts.

The broader field of computer vision has seen a large growth in object detection, classification, and semantic segmentation models. Object detection and classification consist of identifying the presence of an object in an image and categorizing it from a list of possible classes.  Semantic segmentation involves identifying the portion of an image which belongs to a specific class, i.e. deblending.  Put together, these tasks amount to \textit{instance segmentation}.  This pixel-level masking can be used to deblend objects by selecting the pixels associated with each individual object by class.  The benchmark leader in deep learning instance segmentation models has been the Mask-RCNN framework \citep{he_mask_2018}.  

The Mask R-CNN architecture was implemented in \cite{burke_deblending_2019} to detect, deblend, and classify large scenes of simulated stars and galaxies.  Other architectures have been tested in astronomical contexts, including You Only Look Once \citep[YOLO][]{YOLOv4}. \cite{He21} use a combination of the instance segmentation model YOLOv4 and a separate classification network to perform source detection and classification on SDSS images, and \cite{Gonzalez2018} use a YOLO model to detect and morphologically classify SDSS galaxies.  However, these models do not perform segmentation.

The rapid pace of research has led to many new variations and methods that can outperform benchmark architectures. To the benefit of computer vision researchers, Facebook AI Research (FAIR) has compiled a library of next-gen object detection and segmentation models under the framework titled \textsc{Detectron2} \citep{wu2019Detectron2}. This modular, fast, and well-documented library makes a fertile testing ground for astronomical survey data. In addition to a variety of architectures, pre-trained models are also provided. By leveraging \textit{transfer learning}, i.e., the transfer of a neural network's knowledge from one domain to another, we can cut back on training time and costs with these pre-trained models. It is also possible to interface new models with \textsc{Detectron2}, e.g., \cite{Li21, maskformer21}, taking advantage of its modular nature and flexibility\footnote{See \hyperlink{https://github.com/facebookresearch/Detectron2/tree/main/projects}{https://github.com/facebookresearch/Detectron2/tree/main/projects} for a comprehensive list of projects.}. 

In this work, we leverage the resources of the \textsc{Detectron2} library by testing state-of-the-art instance segmentation models on large scenes, each containing hundreds of objects.  We perform object detection, segmentation, and classification simultaneously on large multi-band HSC coadds.  Many deep learning applications have been tested on simulated images, but methods applied to real data are often limited by a lack of ground truth. Here, we construct a methodology for using instance segmentation models on real astronomical data, and demonstrate the potential and challenges of this framework when applied to deep images. The HSC data is ideal for testing this framework, as it represents the state-of-the-art among wide/deep surveys, and is closest in quality to upcoming LSST data.  By interfacing with \textsc{Detectron2}, we are able to test new models as the repository is updated.  We compare models with different performance metrics, and test how robust they are to contrast scalings that alter the dynamic range of the data, which will be important to consider for application to other datasets.

The major contributions of this work can be summarized as 1) Using instance segmentation models to deblend and classify objects in real images from HSC.  This demonstrates the feasibility for future integration with wide/deep survey pipelines.  We will show that the models can learn inherent features in the data that lead to classification performance gains above traditional morphological methods.  2) Comparing the performances of different models when the input data undergoes different contrast scalings. There is no standard method for scaling image data in astronomical studies that use deep neural networks, so we apply a variety of pre-processing scalings to the data for each model.  Dynamic ranges can vary significantly across datasets, and raw data may not be ideal for feature extraction. We test sensitivity to contrast scalings to identify models that will be more easily adapted to different datasets. 3) Interfacing our pipeline with the \textsc{Detectron2} framework to test state-of-the-art models. Of particular note are our tests using transformer-based architectures, an emerging framework in computer vision studies. We will show that these architectures are more robust and accurate than traditional convolutional neural networks in both deblending and classifying objects in large scenes.

This paper is organized as follows. In \S\ref{sec:Detectron2}, we present an overview of \textsc{Detectron2} in which we highlight the flexibility of its modular nature and describe the portion of the available deep learning models we implemented. In \S\ref{sec:implementation}, we describe the curation of our datasets, production of ground truth labels, data preparation and our training procedure.  In \S\ref{sec:results} we present the results of training our suite of models and assess performance with different metrics.  \S\ref{sec:discussion}, we discuss the differences in model capabilities, compare the performance of our pipeline to existing results, and discuss the benefits and drawbacks of our method. In \S\ref{sec:conclusions}, we contextualize our findings and conclude.

\section{Detectron2 Framework}\label{sec:Detectron2}

\begin{figure}
    \centering
    \includegraphics[width=\columnwidth]{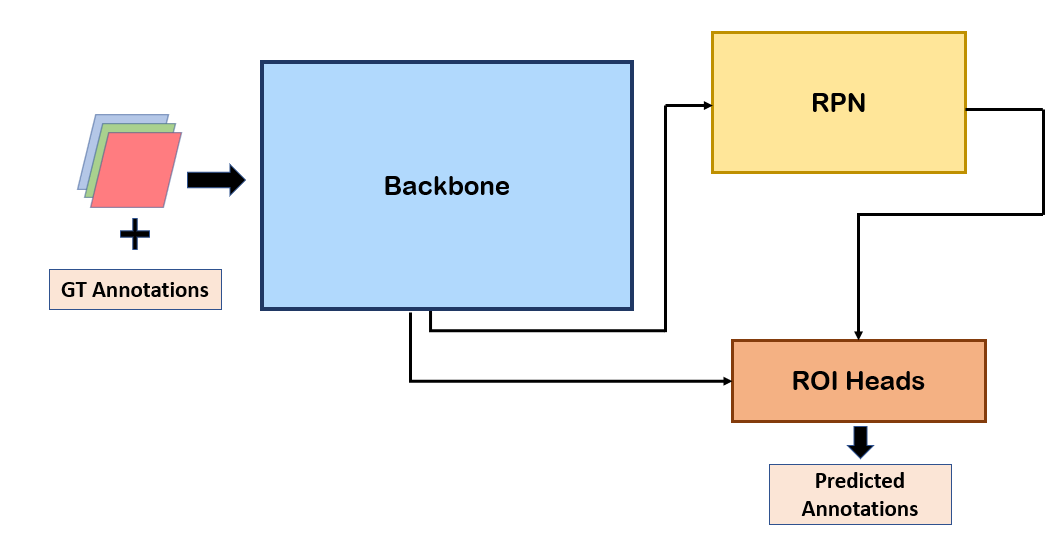}
    \caption{Generalized RCNN meta-architecture.  A multi-channel image along with ground truth object annotations is fed to the backbone feature extractor. These features are passed to the RPN and ROI heads to predict object locations and annotations.}
    \label{fig:detectron_meta}
\end{figure}

We leverage the modular power of \textsc{Detectron2} by implementing models with varying architectures. The pre-trained models we test in \textsc{Detectron2}'s Model Zoo have a structure that follows the GeneralizedRCNN meta-architecture provided by the codebase. This architecture is a flexible overarching structure that allows for a variety of changes, provided they support the following components: (1) a per-image feature extraction backbone, (2) region-proposal generation, (3) per-region feature extraction/prediction.  The schematic of this meta-architecture is shown in Figure \ref{fig:detectron_meta}.

The feature extraction backbone takes an input image and outputs ``feature maps'' by running the input through a neural network, often composed of convolutional layers.  In our tests, we use ResNet backbones and transformer-based backbones.  ResNets are convolutional neural networks that utilize \textit{skip connections} that allow for deep architectures with many layers without suffering from the degrading accuracy problem known to plague deep neural networks \citep{He2016}.  In this paper we explore a few different ResNet backbones: ResNet50, ResNet101 and ResNeXt. A ResNet50 network consists of 50 total layers, with two at the head or ''stem'' of the network and then four stages consisting of 3, 4, 6 and 3 convolutional layers, respectively. Each stage includes a skip connection.  A ResNet101 network is similar to a ResNet50 setup, but with each stage consisting of 3, 4, 23 and 3 convolutional layers, respectively.  Subsequent layers undergo a pooling operation that reduces the input resolution.  We refer the reader to \cite{He2016} for details regarding these layers. ResNeXt layers work similar to ResNet layers, but include grouped convolutions which add an extra parallel set of transforms \citep{Xie16}.  We also test a network with deformable convolutions, in which the regularly spaced convolutional kernel is deformed by a pixel-wise offset that is learned by the network \citep{Dai17}.  

The stages of a ResNet backbone produce feature maps, representing higher level image aspects such as edges and corners.  While one can simply take the feature map outputted by the last layer of the backbone, this can pose a challenge in detecting objects of different scales.  This motivates the extraction of features at different backbones stages (and thus scale sizes).  A hierarchical feature extractor known as a \textit{feature pyramid network} \citep[FPN][]{Lin17} has seen great success in object detection benchmarks. The FPN allows each feature map extracted by a ResNet stage to share information with other feature maps of different scales before ultimately passing on to the Region Proposal Network (RPN). 

After the image features have been extracted, the next stage of Generalized-RCNN networks involves region proposal. This stage involves placing bounding boxes at points in the feature maps and sampling from the proposed boxes to curate a selection of possible objects.  After this sampling has been done, bounding boxes are once again proposed and sent to the Region of Interest (ROI) heads, where they are compared to the ground truth annotations. The annotations consist of bounding box coordinates, segmentation masks, and other information such as class labels.  Ultimately, many tasks can be done on the objects inside these regions of interest, including classification, and with the advent of Mask-RCNN frameworks, semantic segmentation.  We do not include the details of the RPN and ROI heads, as these structures largely remain the same in our tests. We do test architectures with a cascade structure \citep{Cai17} which involves iterating the RPN at successively higher detection thresholds to produce better guesses for object locations. For specifics, we refer the reader to \cite{Girshick2015}, \cite{he_mask_2018} and the \textsc{Detectron2} codebase.

We train a suite of networks to allow for several comparisons. We use a shorthand to denote network configurations as follows.  
\begin{itemize}
  \item R101c4: A ResNet50 backbone that uses features from the last residual stage
  \item R101fpn: A ResNet101 backbone that uses a FPN
  \item R101dc5: A ResNet101 backbone that uses a FPN with the stride of the last block layer reduced by a factor of two and the dilation increased by a factor of two
  \item R50def: A ResNet50 backbone that uses a FPN and deformable convolutions
  \item R50cas: A ResNet50 backbone that uses a cascaded FPN 
  \item X101fpn: A ResNeXt101 backbone that uses a FPN
\end{itemize}
In addition to these ResNet based models, we also test transformer based architectures. A transformer is a encoder-decoder model that employs \textit{self-attention}.  Briefly, self-attention consists of applying linear operations to an encoded sequence to produce intermediate ``query, key and value'' tensors. A further series of linear operations and scalings are done to these intermediate tensors to produce an output sequence, and then a final linear operation is performed on the entire output sequence.  Transformer models have exploded in popularity in the domain of natural language processing due to their scalability and generalizability on sequences, which translates well to language structure.  Recently, transformers have been used in computer vision tasks such as image classification and object detection.  These models been shown to be competitive with the dominant convolutional neural networks, and are seeing rapid advances in performance measures \citep{Dosovitskiy20, DINOv1, DINOv2, Liu21, Li21}.  For example, MViTv2 utilizes multi-head pooling attention \citep[MHPA][]{Fan21} to apply self-attention at different image scales, allowing for the detection of features of varying sizes. To obtain the input encoded sequences, an image is first divided into patches which are flattened and sent through a linear layer.  MHPA is applied to the sequences to produce the image features. In an object detection context, these features are input to an FPN in the same way as features obtained from a ResNet in RCNN models.  Another modern transformer model, the Swin Transformer \citep{Liu21}, applies multi-head attention to image patches, but rather than a pooling operation, use patch merging to combine features of different image patches.  Swin models also use shifted window attention to allow for efficient computation and information propagation across the image. We test both MViTv2 and Swin backbones in our implementation.

\section{Implementation}\label{sec:implementation}

\subsection{HSC coadds}
\label{subsec:HSC_coadds}
In this work, the data we use consist of multi-band image coadds of roughly 4000 pixels$^2$ from the Deep and Ultra-Deep fields of the Hyper Suprime Cam (HSC) Subaru Strategic Program \citep[SSP;][]{Aihara2018} Data Release 3 \citep{aihara_third_2022}. The HSC SSP is a three-tiered imaging survey using the wide-field imaging camera HSC. The HSC instrument \citep{Miyazaki18} consists of a 1.77 deg$^2$ camera with a pixel scale of 0.168'', attached to the prime focus of the Subaru 8.2 m telescope in Mauna Kea.  The Deep+UltraDeep component of the HSC survey covers $\sim$36 deg$^2$ of the sky in five broad optical bands ($grizy$; \citep{Kawanomoto2018}) up to a full 5$\sigma$ depth of $\sim$27 mag (depending on the filter). Despite limitations (e.g., sky subtraction and crowded field issues), the HSC DR3 data provides the closest match among all currently available deep-wide surveys to the expected data quality of LSST wide fields. The Deep/Ultra-Deep field properties are listed in Table \ref{tab:hsc_prop}.  We use the \textit{g}, \textit{r} and \textit{i} bands.
\begin{table}
    \centering
    \begin{tabular}{cccc}
        \hline
        \hline
         & median exposure (min) & seeing (``) & depth (mag) \\
         \hline
         g &  70 & 0.83 & 27.4 \\
         r &  66 & 0.77 & 27.1 \\
         i &  98 & 0.66 & 26.9 \\
         \hline
    \end{tabular}
    \centering
    \caption{Properties of the HSC Deep/UltraDeep images}
    \label{tab:hsc_prop}
\end{table}

Given the large depth of the survey, a significant portion of objects are blended in comparison to other ground-based surveys such as the Dark Energy Survey \citep{DES16}. 
For reference, 58\% of objects in the the shallowest field (Wide) of the HSC survey are blended. While a significant challenge, this lends the HSC fields to be an excellent set of data for testing deblending algorithms, particularly those suited for crowded fields. The pipeline to produce the image coadds is described in detail in \cite{bosch_hyper_2018}. 
There are two sets of sky-subtracted coadds.  The first set consists of global sky-subtracted coadds. The second set also uses the global sky-subtracted images, but an additional local sky subtraction algorithm is applied.  This is to remove the wings of bright objects, artifacts that can cause problems in object detection algorithms. However, this process creates a trade-off with removing flux from extended objects, and \cite{aihara_hyper_2018} empirically find a local sky subtraction scale of 21.5 arcseconds to be a good balance. Ultimately, we use these local sky-subtracted images, as bright wings and artifacts can introduce problems of over-deblending or ``shredding'' and we want our ``ground truth'' detections to be as clean and accurate as possible. 
To further ensure a clean training set, we apply a few quality cuts to the sample.  Some images suffer from missing data in one or more bands, especially at the edge of the imaging fields. We use the bitmasks provided in the coadd FITS files to exclude images with >30\% of the pixels assigned a NO$\_$DATA flag. Given that the neural network takes multi-band images, if one of the \textit{g}, \textit{r} or \textit{i} band images is flagged in this way, we exclude the other bands as well. There remain some imaging artifacts and issues, such as saturated regions around bright stars, and we discuss how these affect network performance in Section \ref{subsec:missing_label_bias}.

\subsection{Ground Truth Generation}
\label{sec:ground_truth}

\begin{figure}
    \centering
    \includegraphics[width=\columnwidth]{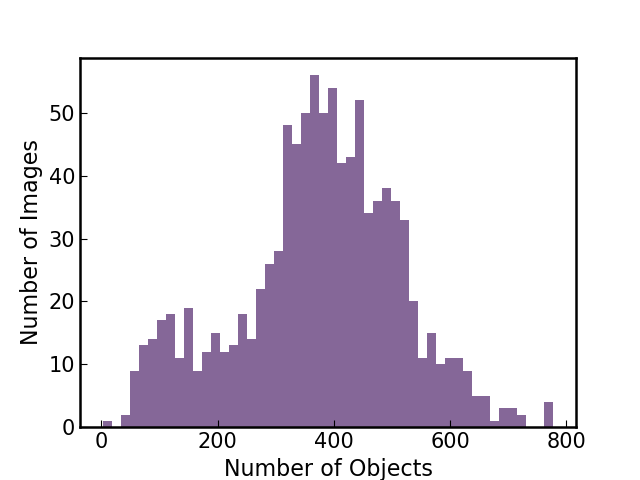}
    \caption{Histogram of the number of objects detected at >5$\sigma$ above the background for HSC images in the training set. The images are taken from both the Deep and UltraDeep fields. }
    \label{fig:hsc_obj_hist}
\end{figure}

We must provide ground-truth object locations and masks to the network to perform pixel-level segmentation. We utilize the multi-band deblending code \textsc{scarlet} \citep{melchior_scarlet_2018} to produce a model for each individual source from which we create an object mask.  \textsc{scarlet} utilizes constrained matrix factorization to produce a spectral decomposition of an object.  It is a non-parametric model that has been demonstrated to work well on individual galaxies and blended scenes. Before we run \textsc{scarlet}, we extract an object catalog using \textsc{sep}, the python wrapper for \textsc{SExtractor}. Then, each identified source is modelled and the ``blend'' or composition of sources is fit to the coadd image data. Once the final blend model is computed, the mask is determined by running \textsc{sep} on each individual model source and setting a mask threshold of 5$\sigma$ above the background. Both the \textsc{scarlet} modelling and mask thresholding are done on the detection image, i.e., the sum over all bands.  The run time of this process increases with the number objects in an image. In order to reduce run-time, we divide the 4k stitched coadd images into 16 images of $\sim$1000$\times$1000 pixels$^2$.  While \textsc{scarlet} on its own is a powerful deblender, the fits can take up to $\sim$30 minutes depending on the number of objects in the image, which motivates the use of efficient neural networks.  After this process is complete, we compile a training set of 1000 1k$\times$1k pixels$^2$ images. The distribution of the number of sources per image is shown in Figure \ref{fig:hsc_obj_hist}.

\begin{figure}
    \includegraphics[width=\columnwidth]{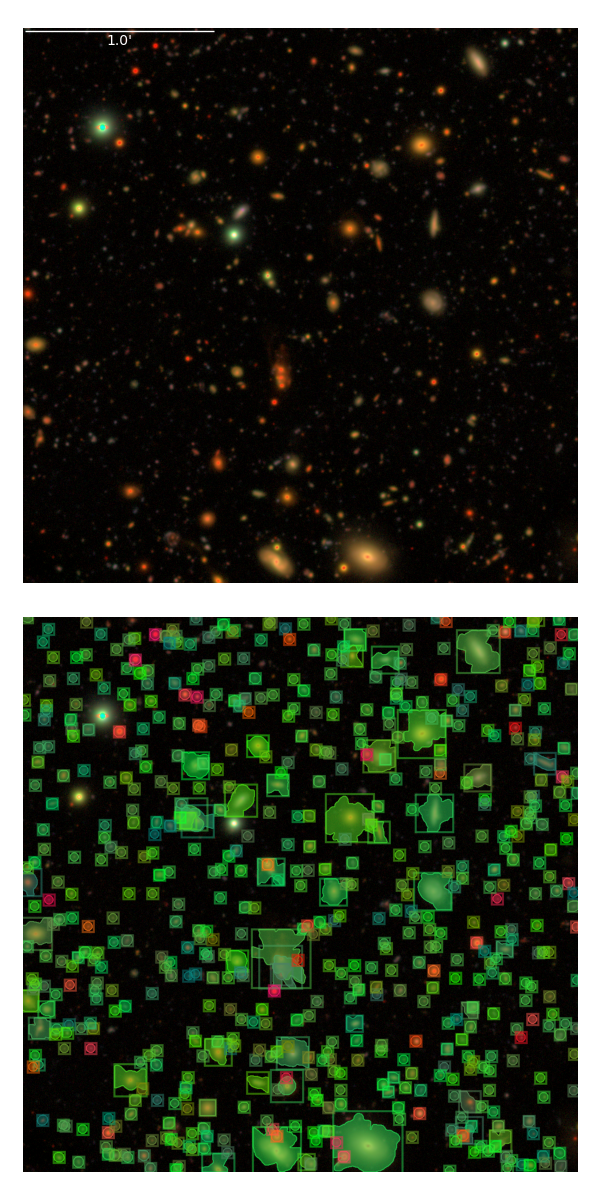}
    \caption{The ground truth masks and bounding boxes on an example image in the test set of our HSC Deep/UltraDeep field data. The image without overlaid masks/boxes in shown below for clarity. A Lupton contrast scaling is used in this visualization. Galaxies are colored green, and stars are colored red.}
    \label{fig:hsc_gt}
\end{figure}

The trade-off in using real over simulated data is that in supervised tasks, there is a lack of predetermined labels. For the classification task, we produce object labels with a catalog match to the HSC DR3 catalogs. We convert each detected source center to RA and DEC coordinates and then run the \textsc{match\_to\_catalog\_sky} algorithm in astropy to find objects in the HSC catalog within 1 arcsecond.  Then, we compare the \textit{i}-band magnitude of the deblended source to the ``cmodel'' magnitude of the catalog objects and pick the object with the smallest magnitude difference.  If no objects are within 1 arcsecond or no objects have a magnitude difference smaller than 1, we discard the object from our labelled set. Once an object is matched, we use the HSC catalog ``extendedness value'' to determine classes, which is based on differences in PSF magnitudes and extended model magnitudes.  While yielding high accuracy at bright magnitudes, this metric becomes unreliable for star classification around a limiting magnitude of 24 mag in the i band \citep{bosch_hyper_2018}. We additionally discard objects with NaN values in the DR3 catalog, as the class is indeterminate.  We show an example image and the results of our labelling methodology in Figure \ref{fig:hsc_gt}, with color-coded classes.  

\subsection{Data Preparation}
\label{sec:standardization}

\begin{figure*}
    \centering
    \begin{subfigure}{\textwidth}
         \centering
         \includegraphics[width=\textwidth]{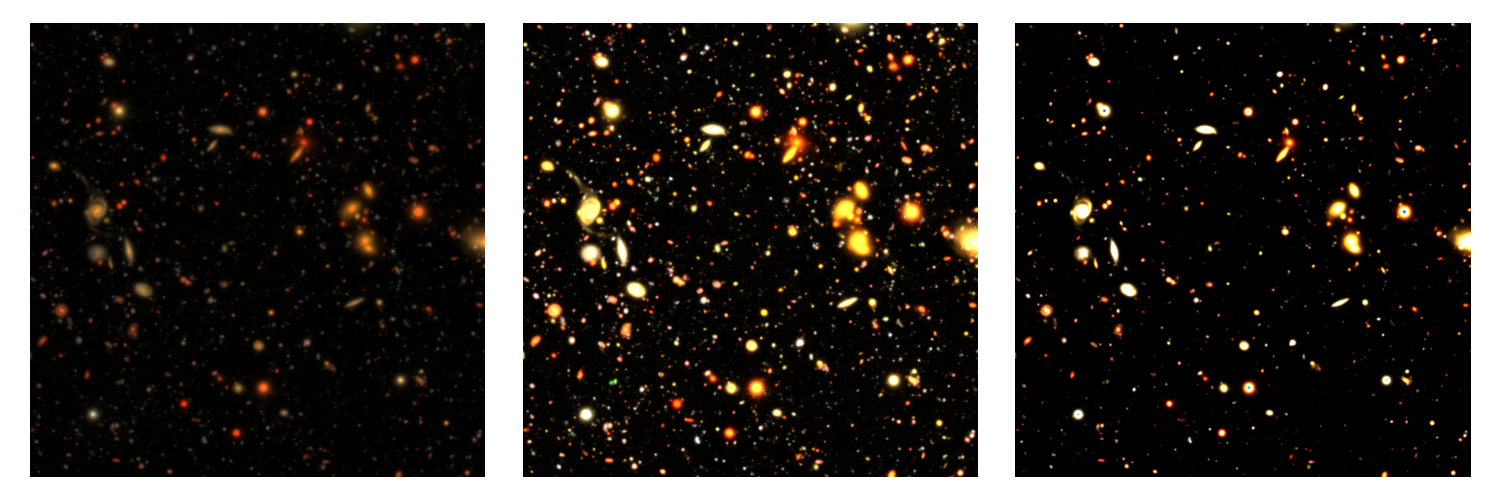}
     \end{subfigure}
    \begin{subfigure}{\textwidth}
         \centering
         \includegraphics[width=\textwidth]{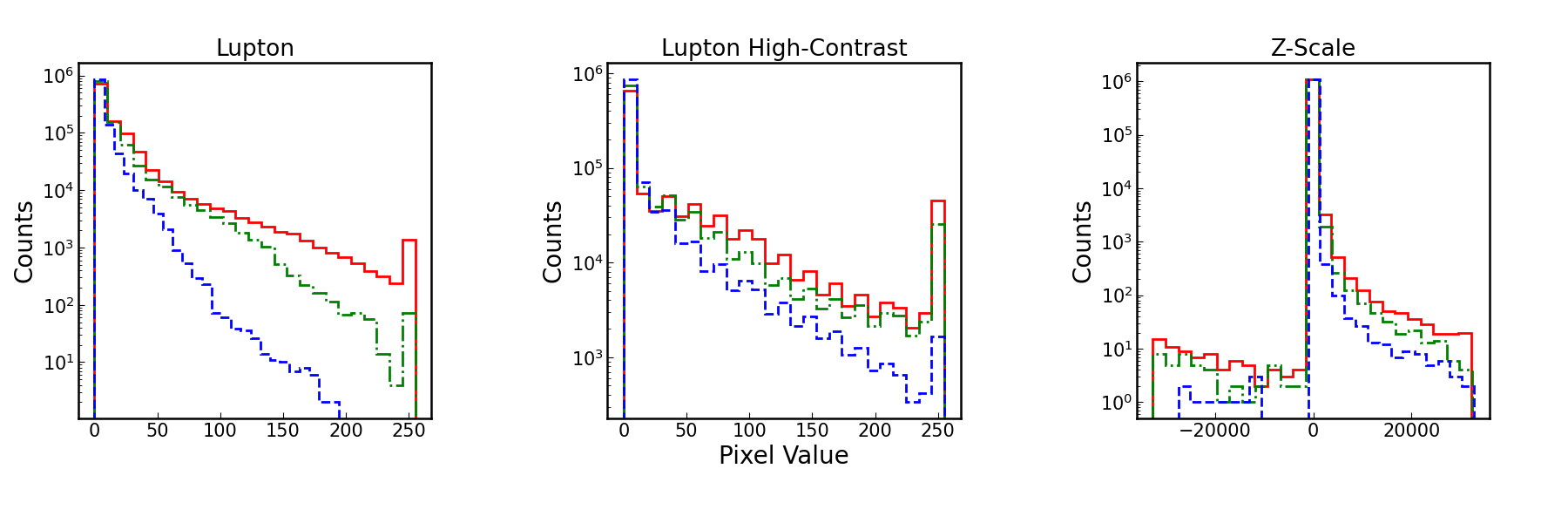}
     \end{subfigure}
    \caption{Top row: RGB images in the HSC DR3 dataset with different contrast scalings.  The scalings are, from left to right: Lupton, Lupton high contrast, and z-scale.  Bottom: Histograms of pixel values to the corresponding image in the top row. Red, green, and blue represent values in the \textit{i}, \textit{r}, and \textit{g} filters, respectively.}
    \label{fig:pixel_scalings}

\end{figure*}

We employ three common methods for scaling the raw data from the coadd FITS files to RGB values.  These are: a z-scale, a Lupton scale, and a high-contrast Lupton scale. The z-scale transformations are commonly employed in computer vision tasks and are given by 
\begin{equation}
    \label{eq:norm}
    \begin{aligned}
    R=A(i-\bar{I})/\sigma_I \\
    G=A(r-\bar{I})/\sigma_I \\
    B=A(g-\bar{I})/\sigma_I \\
    \end{aligned}
\end{equation}
where $I=(i+r+g)/3$ with a mean $\bar{I}$ and standard deviation $\sigma_I$,  $R$ is pixel values in the red channel (and similarly for the green $G$ and blue $B$ channels using the $r$ and $g$ -bands respectively).  We set $A=10^{3}$ for the training and cast the images to 16-bit integers. In addition to z-scaling, we also apply a Lupton scaling from \cite{Lupton2004}. This is an asinh scaling with
\begin{equation}
    \label{eq:norm_asinh}
    \begin{aligned}
    R=i(\textrm{asinh}(Q (I - \textrm{minimum})/\textrm{stretch})/Q \\
    G=r(\textrm{asinh}(Q (I - \textrm{minimum})/\textrm{stretch})/Q \\
    B=g(\textrm{asinh}(Q (I - \textrm{minimum})/\textrm{stretch})/Q .\\
    \end{aligned}
\end{equation}
We use a stretch of 0.5 and $Q=10$ and set the minimum to zero and cast the images to unsigned 8-bit integers. Lupton scaling brings out the fainter extended parts of galaxies while avoiding saturation in the bright central regions. These augmentations preserve the color information of objects to aid in classification.  Lastly, we also use a high-contrast Lupton scaling, in which image brightness and contrast is doubled after applying the Lupton scaling.  We test all of these scalings for each network architecture.  In Figure \ref{fig:pixel_scalings}, we show an example image and a histogram of pixel values in \textit{i}, \textit{r} and \textit{g} bands (corresponding to RGB colors)  

We apply data augmentation to the training and test sets.  Data augmentation has become a staple of many deep learning methods. It allows the network to ``see'' more information without needing to store extra images in memory.  We employ spatial augmentations of random flips and 90$^{\circ}$ rotations.  We do not employ blurring or noise addition, as the real data we train on is already convolved with a PSF and contains noise.  For future generalizations of this framework to different datasets, then blur/noise augmentations may be useful, but for inference purposes on test data taken under the same conditions as the training data, spatial augmentations are sufficient.  We also employ a random 50\% crop on each image during training so that the data can fit into GPU memory.  We considered applying all contrast scalings as a data augmentation, but did not find a significant improvement in network performance.  However, this could be used in future work to reduce the training costs, as results were on par with networks trained with only one contrast scaling.

\subsection{Training}
\label{sec:training}
Training is done using stochastic gradient descent to update the network weights by minimizing a loss function.  The loss functions of these Mask-RCNN models is
\begin{equation}
    L = L_\text{cls} + L_\text{box} + L_\text{mask}
\end{equation}
where the classification loss $L_\text{cls}$ is $-\log p_u$ or the log of the estimated probability of an object belonging to its true class $u$. Discrete probability distributions are calculated per class (plus a background class) for each ROI. $L_\text{box}$ is a smoothed L1 loss calculated over the predicted and true bounding box coordinates as given in \cite{Girshick2015}.  Finally, the mask loss $L_\text{mask}$ is the per-pixel average binary cross-entropy loss between the ground truth and predicted masks.  

All networks are pre-trained on either the MS-COCO \citep{Lin2014} or ImageNet-1k \citep{ImageNet} datasets of terrestrial images, and so we use transfer learning to apply these models to the our astronomical datasets. Transfer learning \citep{Tan2018} is a technique in deep learning that allows for a network trained to do a task in a source domain to perform the task in a different target domain.
%generalize knowledge of one task to complete a different but related task (See \citealt[]{Tan2018} for an overview of deep transfer learning) 
It is often used when applying a pre-trained deep learning model to a different domain than the one seen during training. By using pre-trained weights as initial conditions, training is likely to converge faster and be less prone to over-fitting. We use weights provided by \textsc{Detectron2} as the starting point for our training procedure. We then train the networks for 50 total epochs, i.e. the entire training set is seen 50 total times by the network.  In order to facilitate the transfer of knowledge, we first freeze the feature extraction backbones of the models and only train the head layers in the ROI and RPN networks for 15 epochs. We use a learning rate of 0.001 for this step. Then, we unfreeze the feature extraction backbone and train the entire network for 35 epochs.  We begin this step with a learning rate of 0.0001 and decrease by a factor of 10 every 10 epochs.  To see if transfer learning introduced a bias from terrestrial image pre-training, we trained a model from scratch with randomized weights and compared to a model initialized with pre-trained weights, using the same trainign schedule. We found the pre-trained model to yield the best results, indicating that this step is indeed helpful.

We use two NVIDIA Tesla V100 GPUs in HAL system \citep{10.1145/3311790.3396649} to train on 1,000 images of size 500 pixels$^2$ paired with object annotations. When trained in parallel on each GPU, our models take roughly $\sim$3 hours to complete. Transformer architectures tend to use more memory, and thus are trained on 4 GPUs for roughly 4 hours. 

\section{HSC Results}
\label{sec:results}

After training, we evaluate network performance on the test set of HSC images.  The test set is taken from the patches in the UltraDeep COSMOS \citep{COSMOS} field and consists of 95 images of 1000 pixels$^2$. No test set images were seen during training.  A benefit of the instance segmentation models used in this work is their ability to infer on images of variable size.  Thus, despite the need to crop images during training, we are still able to utilize the full size of the images in the test set. 

\begin{figure}
     \includegraphics[width=\columnwidth]{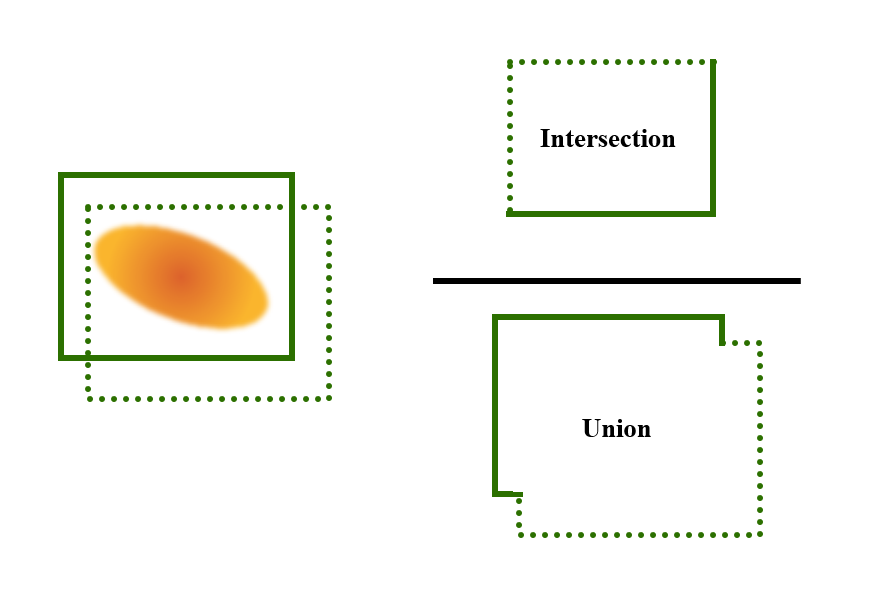}
     \caption{Intersection Over Union (IOU) scores are calculated between a ground truth bounding box (solid lines) and inferred bounding box (dotted line) in the test set of images.  An IOU score can range between 0 (no overlap) and 1 (perfect overlap). This can also be done with object segmentation maps instead of bounding boxes.}
     \label{fig:iou_ex}
\end{figure}

We utilize two object classes, galaxy and star, and evaluate classification performance with precision and recall, given by

\begin{equation}
    p=\frac{\text{TP}}{\text{TP}+\text{FP}},
    \label{eq:precision}
\end{equation}
\begin{equation}
    r=\frac{\text{TP}}{\text{TP}+\text{FN}}.
    \label{eq:recall}
\end{equation}
True positives (TP) are counted as a detection that has a confidence score outputted by the network above a certain threshold and additionally can be matched to a ground truth object by having an Intersection over Union (IOU) above another threshold.  Figure \ref{fig:iou_ex} shows an example of how the IOU is calculated for a pair of objects. False negatives (FN) are those ground truth objects that do not have a corresponding detection.  False Positives (FP) are those detections with a high confidence score but do not have a matching ground truth. The IOU is defined as
\begin{equation}
    \text{IOU}=\frac{\text{\emph{area}}(\text{box}_\text{predicted} \cap \text{box}_\text{truth})}{\text{\emph{area}}(\text{box}_\text{predicted} \cup \text{box}_\text{truth})}.
    \label{eq:IOU}
\end{equation}
or the area of the intersection over the area of the union of the predicted and ground truth bounding boxes.  
Precision and recall are often broken down by class, or combined into one value, the AP score,
\begin{equation}
    \text{AP}=\frac{1}{N_\text{thresh}}\sum_{r\in\{0,0.02,...,1.0\}} p(r)
\end{equation}
where $p(r)$ is the maximum precision in a recall bin of width $\Delta r$. AP scores are computed over N$_\text{thresh}$ equally spaced IOU thresholds from 0.5-0.95 and averaged.  Here, N$_\text{thresh}=51$.
%for IOU thresholds of \{0.5,0.55...0.95\} and averaged.
\begin{table*}
\centering
\begin{tabular}{cccccccc|cc}
\hline
\hline

\multicolumn{2}{c}{} &
\multicolumn{6}{c}{ResNets} &
\multicolumn{2}{c}{Transformers} 
\\ \\
 &  & R101C4 & R101dc5 & R101fpn & R50cas & R50def & X101fpn & MViTv2 & Swin \\  
\hline
\multirow[t]{3}{*}{Galaxies} & Lupton & 23.7 & 24.6 & 40.9 & 46.3 & 41.7 & 41.4 & \textbf{51.7} & 50.8 \\
 & LuptonHC & 26.1 & 28.0 & 43.6 & 46.0 & 43.2 & 43.1 & \textbf{50.9} & 50.3 \\
 & zscale & 22.9 & 30.7 & 40.2 & 39.6 & 21.8 & 34.1 & \textbf{52.7} & 52.5 \\
 \hline
\multirow[t]{3}{*}{Stars} & Lupton & 10.3 & 9.6 & 7.3 & 7.4 & 4.3 & 2.5 & \textbf{34.1} & 33.9\\
 & LuptonHC & 2.4 & 5.1 & 6.1 & 8.1 & 5.5 & 8.3 & \textbf{28.0} & 25.0 \\
 & zscale & 15.6 & 10.5 & 17.9 & 25.5 & 12.7 & 17.2 & \textbf{35.8} & 33.9 \\
 \hline
\multirow[t]{3}{*}{Small} & Lupton & 17.6 & 18.0 & 26.1 & 28.0 & 24.6 & 23.7 & \textbf{43.7} & 43.1 \\
 & LuptonHC & 14.8 & 17.2 & 25.9 & 27.7 & 25.4 & 26.9  & \textbf{40.1} & 38.4 \\
 & zscale & 19.7 & 21.5 & 30.2 & 33.2 & 18.1 & 26.8 & \textbf{44.8} & 43.8 \\
 \hline
\multirow[t]{3}{*}{Medium} & Lupton & 8.7 & 11.9 & 14.4 & 11.5 & 13.7 & 11.7 & \textbf{17.4} & 16.1 \\
 & LuptonHC & 7.8 & 11.1 & 13.4 & 12.7 & 10.3 & 12.6 & \textbf{16.3} & 15.5 \\
 & zscale & 3.8 & 9.0 & 7.2 & 7.3 & 1.6 & 3.6 & \textbf{15.1} & 14.9 \\
 \hline
\multirow[t]{3}{*}{Large} & Lupton & 16.4 & 30.9 & 18.9 & 14.3 & 19.6 & 9.3 & \textbf{43.1} & 41.5\\
 & LuptonHC & 15.3 & 22.8 & 14.9 & 15.0 & 11.6 & 13.0 & 38.6 & \textbf{39.7} \\
 & zscale & 0.7 & 3.6 & 3.8 & 5.2 & 0.1 & 0.9 & \textbf{37.8} & 37.0 \\
 \hline
\end{tabular}
\caption{AP scores on COSMOS HSC set for all network configurations (larger is better).  Galaxy and Star AP scores are calculated separately, whereas Small (0-32 pixels$^2$), Medium (32-96 pixels$^2$) and Large (>96 pixels$^2$) object AP scores are averaged across both classes. The best result for each row is emphasized in bold. The MViTv2 backbone gives the best results in all cases except for one.}
\label{tab:AP_hsc}
\end{table*}

AP scores on the HSC COSMOS test set are reported for all network configurations in Table \ref{tab:AP_hsc}.  We report the per-class AP score for stars and galaxies separately, as well as the Small, Medium, and Large AP scores, defined by the object bounding box size of 0-32 pixels$^2$, 32-96 pixels$^2$ and >96 pixels$^2$, respectively. For galaxies and stars, AP score can vary significantly across network configurations.  For ResNet-based architectures, AP for galaxies is consistently higher than stars, which may be due to the higher sample size of galaxies and morphological features that make galaxies easier to distinguish than compact stars.  Among ResNet-based networks, a Lupton high-contrast scaling generally gives the highest galaxy AP score, while a z-scaling always gives the highest star AP score.  It appears that these networks are very sensitive to the contrast scaling used, which is not desirable for application to other datasets with different dynamic ranges.  However, transformer-based architectures perform more robustly with varying contrast scalings, and outperform ResNet architectures in almost all cases.  For these networks, galaxy AP scores all lie within $\sim$50-52, showing a gain of about 5 over the highest performing ResNet configuration.  Stellar AP scores for Lupton and z-scalings lie within $\sim$33-35, with high-contrast Lupton scalings performing worse by an AP of $\sim$8.  
Among the Small, Medium, and Large AP metrics, transformer-based networks also outperform ResNet-based networks, in some cases seeing massive gains in AP score.  The networks generally perform better on Small and Large object categories over Medium objects, again likely due to sample size.  

Many studies of instance segmentation models use the MS-COCO or ImageNet-1k datasets as a benchmark to judge performance through the AP score.  These data consist of terrestrial images with many object classes, so it can not necessarily be used as a comparison for our AP scores calculated on astronomical survey images with only 2 classes.  However, to give a reader a sense of the range of typical values, the AP scores for models trained on terrestrial data typically range from $\sim$35-45 for convolutional backbones and push to $\sim$55 for transformer backbones (see the \textsc{Detectron2} repo for results).  For a more fair comparison, we look to \cite{burke_deblending_2019} in which instance segmentation models were tested on the simulated observations from the Dark Energy Camera \citep[DECam][]{DECam}.  The authors report an AP score for galaxies of 49.6 and score of 48.6 for stars, averaged to a combined score of 49.0.  We also train our suite of models on the DECam dataset and report the results in Appendix \ref{app:decam}.  More recently, \cite{He21} use a combination of the instance segmentation model YOLOv4 \citep{YOLOv4} and a separate classification network to perform source detection and classification on SDSS images.  They report an AP score of 52.81 for their single-class detection network.

\subsection{Incorrect Label Bias Mitigation}

There is an inherent bias in our measure of AP scores due to incorrect object class labels. In measurements described above, we test the network abilities to infer classes based on labels generated from HSC catalogs.  However, these labels are known to become unreliable, especially for stars, around \textit{i}-band magnitudes of $\sim$24 mag \citep{bosch_hyper_2018}. We use HSC coadds in the COSMOS field for our test dataset, and attempt to mitigate this mislabelling bias by exploiting the overlap of this field with space-based observations using the Advanced Camera for Surveys (ACS) on the Hubble Space Telescope (HST).  Because of the lack of atmospheric seeing, morphological classification of stars/galaxies using the HST COSMOS catalog data is much more precise for faint objects, and can be used as ground truth instead of HSC labels. This will test how much poor classification behaviour is due to label generation as opposed to limitations of the models. We generate HST labels by cross-matching detected sources to the catalog of \cite{Leauthaud07} within 1 arcsecond. If there is no object within 1 arcsecond, we discard the object. There is not necessarily a one-to-one match of HSC versus HST labels, as we are cross-matching to different catalogs, but the number of objects per image remains roughly the same for either labelling scheme. We will refer to this as the HST COSMOS test set.

\begin{figure}
    \centering
    \includegraphics[width=\columnwidth]{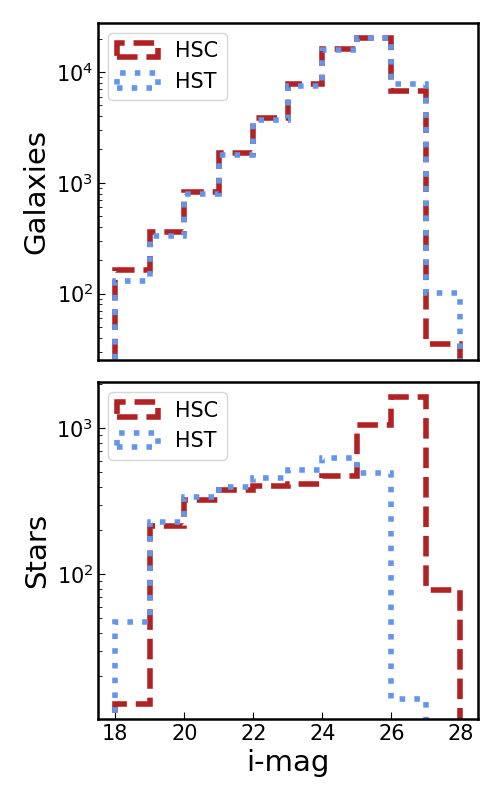}
    \caption{Galaxy and star counts for our COSMOS set, with labels generated from HSC and HST catalogs.  The extra counts of HSC stars at faint magnitudes is due to galaxy contamination when classification is based on the extendedness metric. The low sample of bright HSC stars follows from our catalog matching procedure of excluding objects with NaN values.}
    \label{fig:COSMOS_labels}
\end{figure}

Given the size of the HST/HSC overlap in the COSMOS field and the size of our cutouts, there is not enough coverage to produce a sufficiently large training and test set with HST labels. Instead, we take the models trained on HSC-labelled data and test their evaluation performance on the HST COSMOS test set. To highlight the differences in class label generation, in Figure \ref{fig:COSMOS_labels} we show the number of stars and galaxies as a function of HSC \textit{i}-band magnitude for the COSMOS set for both HSC and HST class labels.  The unreliable quality of HSC labels at faint magnitudes is reflected in the increased counts of stars, especially the bump in stellar counts beginning at \textit{i}$\sim$25 mag.  Also of note is the fewer amounts of star counts in the HSC COSMOS set at bright magnitudes. This is likely due to our HSC label generating procedure of discarding objects with NaN values in the HSC catalog.  Bright stars are likely to have saturated pixels in their centers, causing these error flags to appear.  With HST labels. we can test with a more astrophysically accurate baseline.  

Using this new test set, we present AP scores in Table \ref{tab:AP_hst}.
\begin{table*}
\centering
\begin{tabular}{cccccccc|cc}
\hline
\hline

\multicolumn{2}{c}{} &
\multicolumn{6}{c}{ResNets} &
\multicolumn{2}{c}{Transformers} 
\\ \\
 &  & R101C4 & R101dc5 & R101fpn & R50cas & R50def & X101fpn & MViTv2 & Swin  \\
\hline
\multirow[t]{3}{*}{Galaxies} & Lupton & 25.9 & 26.8 & 42.9 & 49.4 & 43.5 & 42.8 & 51.8 & \textbf{52.4} \\
 & LuptonHC & 27.4 & 30.0 & 46.2 & 50.2 & 46.7 & 44.3  & 51.5 & \textbf{51.6} \\
 & zscale & 25.5 & 32.5 & 42.7 & 41.5 & 23.0 & 35.6 & 52.2 & \textbf{52.9} \\
 \hline
\multirow[t]{3}{*}{Stars} & Lupton & 16.2 & 15.0 & 10.9 & 10.9 & 7.1 & 3.8 & 52.9 & \textbf{53.7} \\
 & LuptonHC & 4.2 & 7.9 & 11.2 & 14.2 & 9.4 & 13.9  & \textbf{42.1} & 37.7\\
 & zscale & 28.3 & 19.1 & 29.3 & 41.6 & 23.8 & 29.0 & \textbf{53.9} & 52.6 \\
 \hline
\multirow[t]{3}{*}{Small} & Lupton & 22.0 & 22.1 & 29.3 & 31.4 & 27.0 & 25.2 & 54.0 & \textbf{54.7} \\
 & LuptonHC & 16.4 & 19.9 & 30.0 & 33.3 & 29.4 & 30.7 & \textbf{48.2} & 46.0 \\
 & zscale & 28.0 & 27.1 & 37.8 & 42.9 & 24.8 & 34.1 & \textbf{54.7} & 54.3 \\
 \hline
\multirow[t]{3}{*}{Medium} & Lupton & 8.3 & 11.7 & 13.8 & 11.0 & 13.1 & 11.1 & \textbf{16.3} & 15.2 \\
 & LuptonHC & 7.5 & 10.8 & 12.7 & 12.2 & 9.9 & 12.0 & \textbf{15.4} & 14.6  \\
 & zscale & 3.7 & 8.5 & 7.3 & 7.4 & 1.7 & 3.6 & \textbf{14.1} & \textbf{14.1}  \\
 \hline
\multirow[t]{3}{*}{Large} & Lupton & 6.2 & 11.1 & 7.2 & 5.9 & 7.2 & 3.6 & \textbf{15.1} & 15.0 \\
 & LuptonHC & 5.4 & 7.9 & 5.3 & 4.8 & 4.4 & 4.8 & 13.7 & \textbf{14.0}  \\
 & zscale & 0.3 & 1.2 & 1.3 & 1.9 & 0.1 & 0.2 & \textbf{13.6} & 13.5 \\
 \hline
\end{tabular}
\caption{Same as Table \ref{tab:AP_hsc}, but with the COSMOS HST test set.}
\label{tab:AP_hst}
\end{table*}
The results for galaxy/star AP scores are in line with the previous results on the HSC COSMOS test set.  In all cases, transformer architectures outperform ResNet architectures and are more robust to different contrast scalings.  AP scores for Small bounding box objects improves for all network configurations, Medium bounding box AP score roughly remains the same, and Large bounding box AP score worsens. The decrease in Large bounding-box AP scores is likely due to the initial label generation step with \textsc{sep} that over-deblends or ``shreds'' large extended galaxies and saturated regions around stars.  With our HSC label generation, we exclude many of the shredded regions by enforcing the i-band $\Delta$1 mag criterion and discarding labels matched to saturated catalog objects with NaN values.  However, our HST label generation is solely based on a distance matching criterion, and so some of these shredded regions are included in the ground truth labels in the HST COSMOS test set.  These spurious extra labels can lead to lower AP scores if the networks avoid shredding these regions at inference.  In the next section, we examine metrics other than AP score that are less susceptible to this effect.

\subsection{Missing and Extra Label Bias Mitigation}
\label{subsec:missing_label_bias}

\begin{figure*}
     \centering
     \begin{subfigure}{\textwidth}
         \centering
         \includegraphics[width=\textwidth]{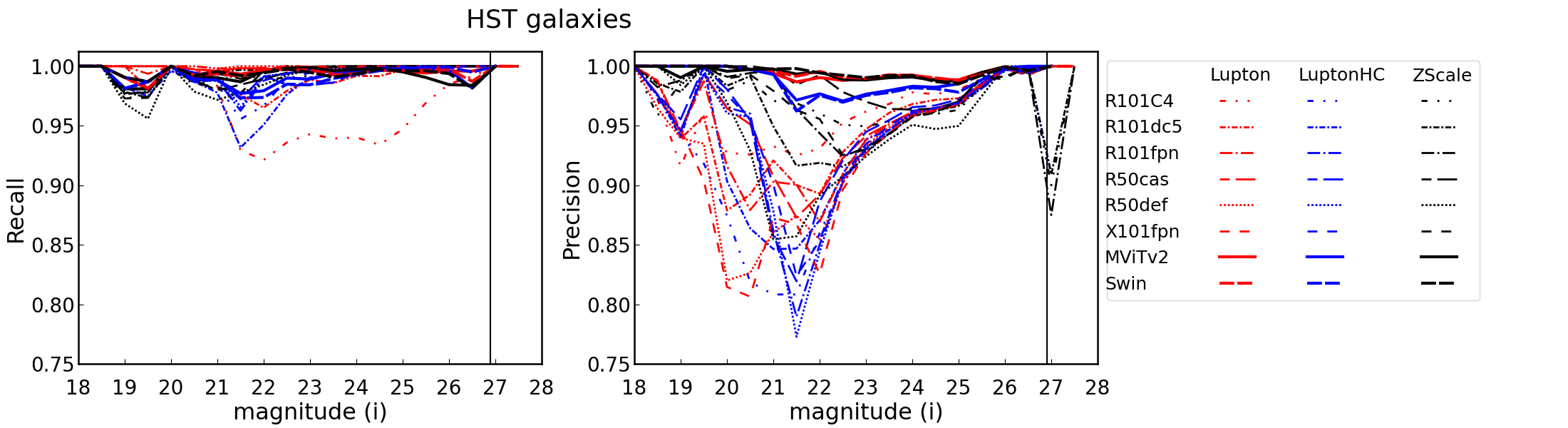}
         \label{fig:y equals x}
     \end{subfigure}
     \hfill
     \begin{subfigure}{\textwidth}
         \centering
         \includegraphics[width=\textwidth]{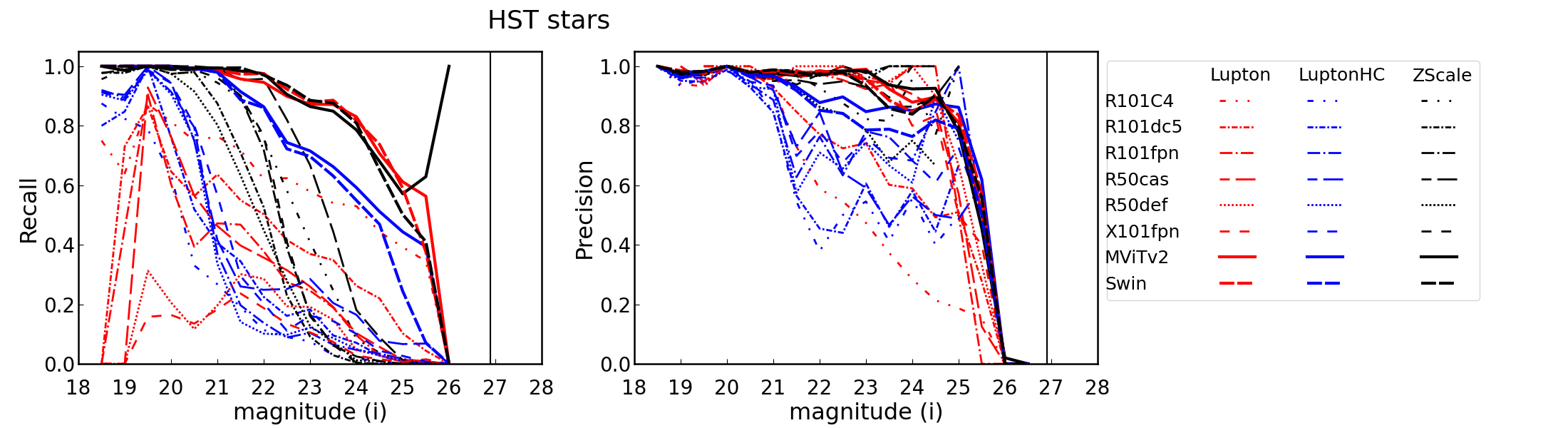}
     \end{subfigure}
     \hfill    
     \caption{Top: Galaxy precision/recall metrics as a function of object magnitude in the HST i-band.  The colors correspond to individual backbone architectures and are shown in the legend.  Linestyles represent different network architectures following the legend, and colors indicate which contrast scaling was used (red for Lupton, blue for LuptonHC and black for z-scale). The black vertical line indicates the Deep/UltraDeep i-band 5$\sigma$ magnitude of 26.9 mag. The y-axis is truncated to better show the differences across the models. Bottom: Stellar precision/recall metrics as a function of object magnitude in the HST i-band. }
     \label{fig:hst_mag_pr}
\end{figure*}

\begin{table*}
\centering
\begin{tabular}{cccccccc|cc}
\hline
\hline
\multicolumn{2}{c}{} &
\multicolumn{6}{c}{ResNets} &
\multicolumn{2}{c}{Transformers} 
\\ \\
 &  & R101C4 & R101dc5 & R101fpn & R50cas & R50def & X101fpn & MViTv2 & Swin \\
\hline
\multirow[t]{3}{*}{Galaxies} & Lupton & 0.963 & 0.981 & 0.982 & 0.983 & 0.980 & 0.979 & \textbf{0.994} & \textbf{0.994} \\
 & LuptonHC & 0.975 & 0.978 & 0.980 & 0.981 & 0.981 & 0.980 & \textbf{0.991} & 0.990 \\
 & zscale & 0.982 & 0.981 & 0.982 & 0.986 & 0.969 & 0.980 & \textbf{0.994} & \textbf{0.994} \\
 \hline
\multirow[t]{3}{*}{Stars} & Lupton& 0.458 & 0.475 & 0.334 & 0.327 & 0.215 & 0.145 & 0.881 & \textbf{0.884} \\
 & LuptonHC & 0.233 & 0.330 & 0.325 & 0.397 & 0.294 & 0.375  & \textbf{0.800} & 0.751 \\
 & zscale & 0.690 & 0.571 & 0.615 & 0.763 & 0.603 & 0.643 & \textbf{0.873} & 0.869  \\
\hline
\end{tabular}
\caption{F1 scores for star and galaxy classes in the HST COSMOS test set, computed for all network configurations. Transformer networks outperform convolutional networks in all cases, especially for stars.}
\label{tab:HST_f1s}
\end{table*}
Since we have done the labelling ourselves using \textsc{sep}, \textsc{scarlet} and catalog matching to produce ground truth detections, masks and classes, traditional metrics of network performance may not be the best choice in characterizing efficacy.  Consider the precision/recall and AP metric.  An implicit assumption in these metrics is the completeness and purity of the ground truth labels. This assumption holds for large annotated sets of terrestrial images such as the MS-COCO set \citep{Lin2014} commonly used as a benchmark in object detection/segmentation studies.  It also holds for simulated datasets of astronomical images \citep{burke_deblending_2019} as the ground truth object locations, masks, and classes are all known \textit{a priori} when constructing the training and test set labels. However, real data of large astronomical scenes presents a challenge. Given that we must generate labels without a known underlying truth, any comparisons to this ``ground truth'' are really comparisons to the methods used to generate these labels. Issues in the label generating procedures will propagate to the performance metrics.  

First, the ground truth detections are produced from running \textsc{sep} using a detection threshold of 5$\sigma$ above the background.  This causes a lack of complete labels, as some objects are missed.  We could lower this threshold, but then run the risk of further over-deblending extended/saturated objects. This leads to the second issue in that there will still remain some level of shredding that will cause spurious extra objects to appear in the ground truth set, i.e, a lack of pure labels.  If the networks do not shred extended/saturated objects as much as \textsc{sep}, (which is a desirable feature of the networks) then the AP metric will be \textit{lower} due to less spurious network detections than the ground truth. Finally, the object detection mechanisms of the neural networks used in this work are fundamentally different from the peak-finding detection used in \textsc{sep}. 

These issues lead to cases in which the neural networks detect objects that are not labelled in our ground truth catalog, despite being actual objects, or cases in which the networks do not detect unphysical objects that are in the ground truth.  Any metric that considers true/false detections is subject to this effect. We do not wish to count these cases of fake true/false positives, as this would lead to a reduction in performance metrics that does not reflect network classification/detection accuracy, but rather the limitations of our label generation.  Therefore, we construct a set of metrics similar to the canonical precision and recall, but slightly alter our definitions of positive and negative detections.  We use equations \ref{eq:precision} and \ref{eq:recall}, but we limit our metrics to the set of objects D that are matched to a ground truth detection. The set of matched detections D is determined by selecting the inferred bounding box with the highest IOU to a ground truth bounding box, above a threshold of 0.5.  Then for a given class C, true positives are the objects in D that are correctly classified, false positives are objects that are incorrectly assigned class C, and false negatives are matched objects with a ground truth class C that the network assigns to a different class.  With these metrics, precision and recall measure purely the classification power of the network, without bias from missing labels or extra false labels. If we assume that the network's ability to classify remains consistent for objects outside of the matched set, we can generalize these metrics to overall classification performance. 

We combine precision and recall into one metric to judge classification power, the F1 score, which is given by the harmonic mean between precision and recall,
\begin{equation}
    \textrm{F}1=2\times \frac{p*r}{p+r}.
    \label{eq:f1}
\end{equation}

The F1 score balances the trade-off between precision and recall, with a value close to unity being desirable.  We report the F1 scores for the networks on the HST COSMOS test set in Table \ref{tab:HST_f1s}. The best performing configuration among ResNet architectures is the R50cas network with a z-scale scaling. A Swin network with a Lupton scaling achieves the highest overall galaxy and star F1 scores, although the MViTv2 architecture remains competitive.  Nearly all transformer networks configurations perform better on star/galaxy classification than ResNet-based networks.  Classification power of transformer-based networks is again more robust to contrast scalings than ResNet-based networks.  

To examine network performance on faint objects, we show precision and recall as a function of \textit{i} band magnitude for the HST COSMOS test set in Figure \ref{fig:hst_mag_pr}. Galaxy recall maintains a value close to one for all objects regardless of magnitude, with some fluctuations of a few percent for some models. Galaxy precision dips for some models at bright magnitudes, which may be due to compact galaxies with bright cores resembling stars.  However, these dips are more likely due to inherent limitations of the models rather than label generation, as transformer architectures produce high galaxy precision and recall across magnitude bins compared to ResNet architectures. Most ResNet architectures suffer with stellar recall, with many showing poor performance even at bright magnitudes. Stellar precision reaches near unity at bright magnitudes for all architectures, but many networks configurations begin to drop in performance around \textit{i} band magnitudes of 21 mag.  The best performing networks maintain a stellar precision above 0.8 out to $\sim$25 mag in the \textit{i} band.  The transformer models we trained are able to achieve a 99.6 percent galaxy recall, 99.2 percent galaxy precision, 85.4 percent stellar recall and 91.5 percent star precision on our HST COSMOS test set, averaged over the whole magnitude range. For comparison, \cite{He21} perform deep neural network object detection and classification of stars, galaxies, and quasars in large SDSS images.  With their sample of objects that covers an \textit{r} band magnitude range of 14-25 mag, they report a galaxy recall of 95.1 percent, galaxy precision of 95.8 percent, stellar recall of 84.6 percent and stellar precision of 94.5 percent.  

\begin{figure*}
	\includegraphics[width=\textwidth]{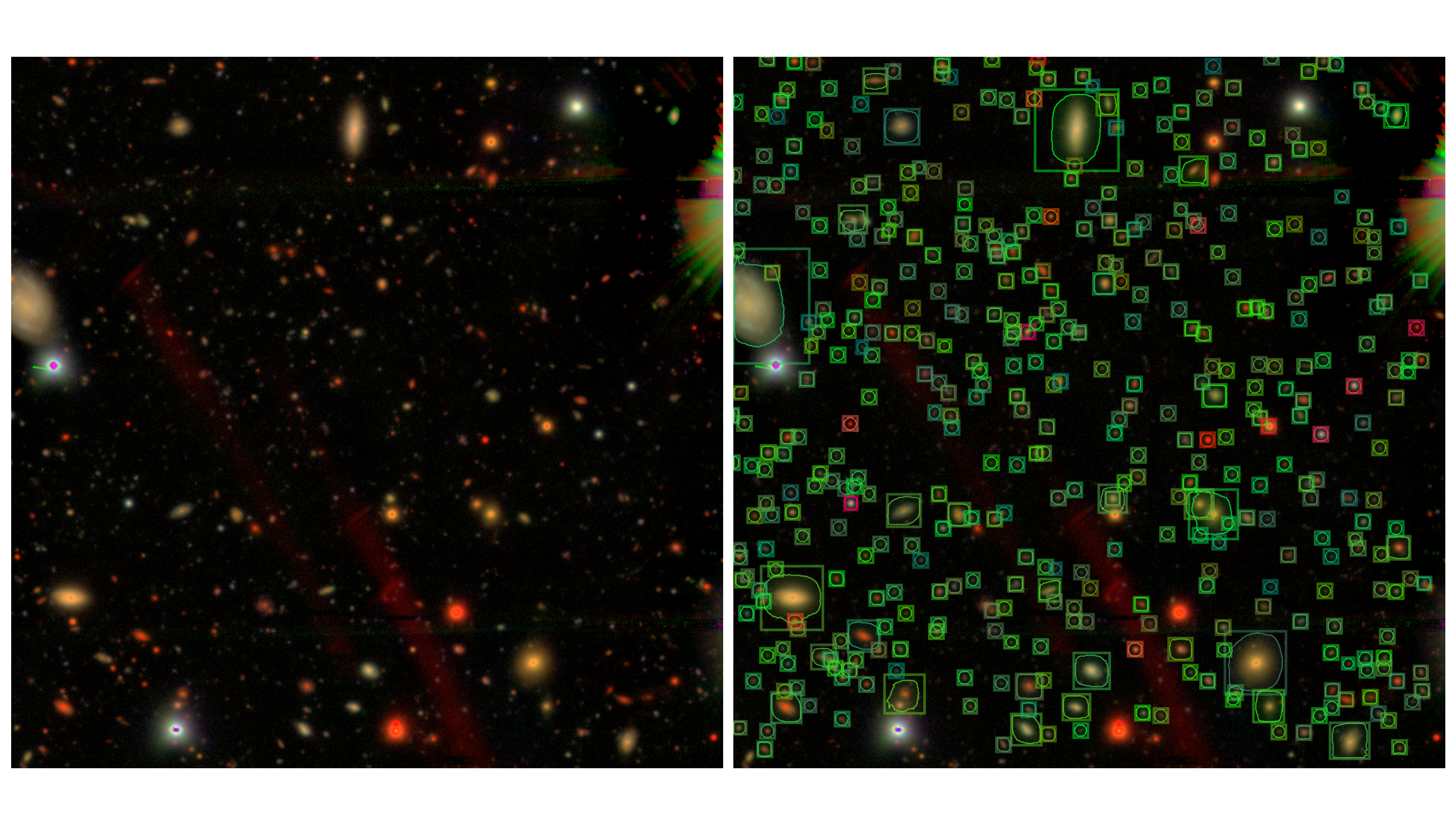}
    \caption{Left: An example image containing multiple artifacts, including blooming and optical ghosts around the bright star in the upper right and large ghosts in the lower middle of the image. Right: The inference results of a MViTv2 Lupton scaled network.  While identifying some objects within the regions of these artifacts, the network does not classify the artifacts themselves as legitimate objects. Bright regions around stars and other artifacts (regardless of location) are generally ignored, as they are not labelled in the training set and thus not "seen" by the networks.}
    \label{fig:mvit_artifacts}
    
\end{figure*}

Some images contain artifacts such as bleed trails, diffraction spikes, or ``ghost images" often appearing around bright stars.  We do not expect the networks to misidentify these artifacts as real objects, as they are largely excluded from our training set due to the label generation procedure described in Section \ref{sec:ground_truth}. Indeed, the networks largely ignore these artifacts, seen for example in Figure \ref{fig:mvit_artifacts}. Explicit identification of artifacts is possible, as in \cite{Tanoglidis22} where the authors use a Mask-RCNN architecture to identify ghosts and other artifacts in DES images.

\subsection{Deblending}
\begin{figure*}
    \centering
    \includegraphics[width=\textwidth]{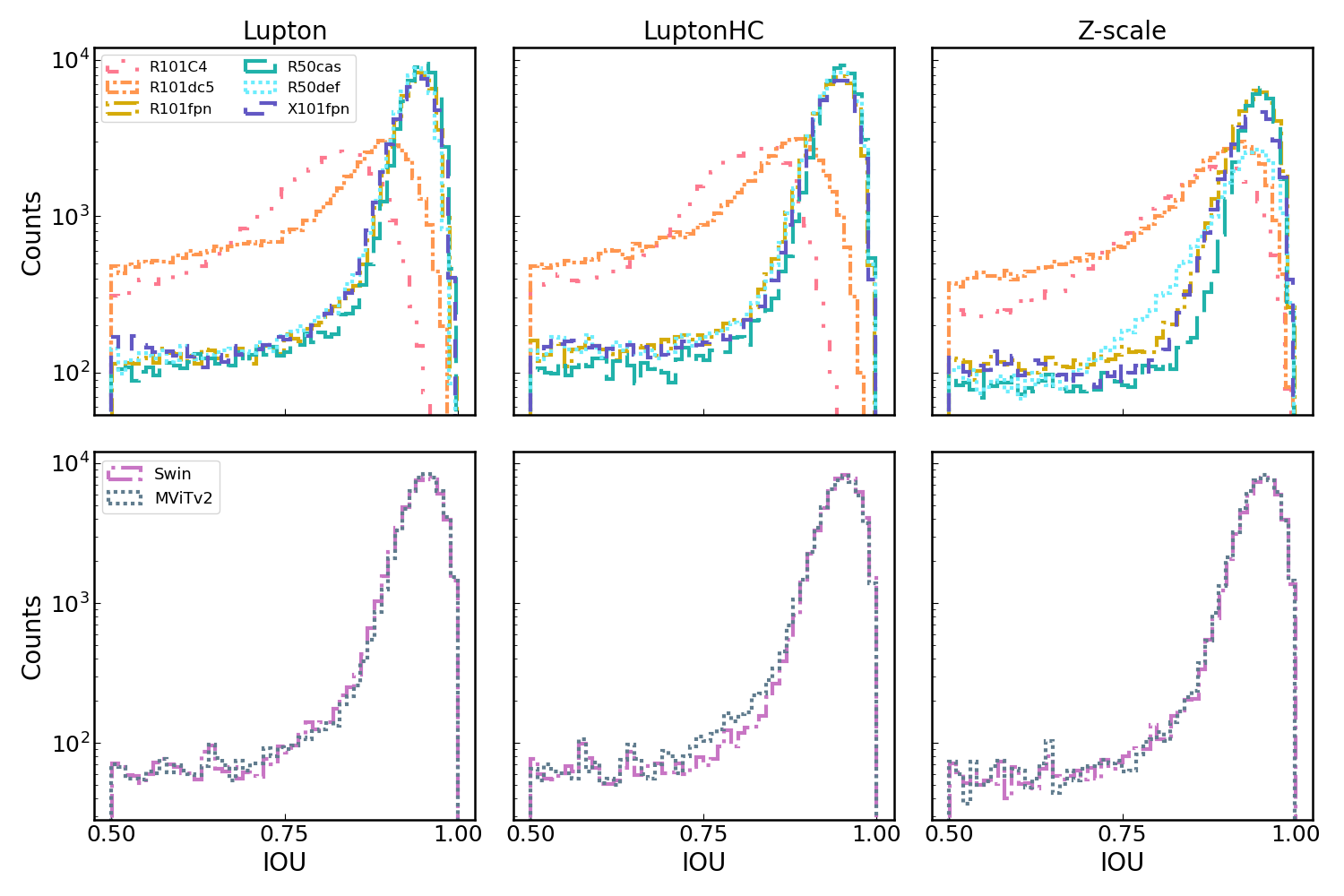}
    \caption{Bounding box IOUs of each detected object that is matched a a ground truth object.  Rows show the results for different transformer backbones. Top: results for ResNet backbones. Bottom: results for transformer backbones. The left column represents Lupton scaling, the middle Lupton high-contrast and the right z-scaling.}
    \label{fig:iou_hist}
\end{figure*}

In order to quantify deblending performance of the networks, we compute IOU scores for matched objects.  The process is similar to the matching done in computing classification precision/recall.  We first set a detection confidence threshold of 0.5 and then compute the bounding box IOUs for all detected and ground truth objects. For each ground truth object, we take the corresponding detected object with the highest IOU above a threshold of 0.5.  We employ this threshold to avoid the biases discussed in Section \ref{subsec:missing_label_bias}.  An IOU of one indicates a perfect match between the ground truth box and the inferred box. In addition to bounding box IOU, we also compute the segmentation mask IOU, which follows from Equation \ref{eq:IOU}, but uses the area of the true and predicted segmentation masks. We report the median IOU for all matched objects in Table \ref{tab:resnet_ious}, and show the distributions in Figure and \ref{fig:iou_hist}.  Transformer-based networks generally produce a higher bounding box IOU than ResNet-based networks, although the R50cas, R101fpn and X101fpn networks remain competitive. Segmentation mask IOUs are lower than bounding box IOUs in all cases.  This indicates that while the networks are able to identify overall object sizes quite well, the finer details of object shapes within the bounding boxes are not as well inferred.  

\begin{table*}
\centering
\begin{tabular}{lllllll|ll}
\hline
\hline

\multicolumn{1}{c}{} &
\multicolumn{6}{c}{ResNets} &
\multicolumn{2}{c}{Transformers}
\\ \\
 & R101C4 & R101dc5 & R101fpn & R50cas & R50def & X101fpn & MViTv2 & Swin\\
\hline
Lup & 0.75 (0.61) & 0.78 (0.57) & 0.93 (0.63) & \textbf{0.94} (0.62) & 0.93 (0.64) & 0.93 (0.64) & \textbf{0.94} (0.64) & \textbf{0.94} (0.64)\\
LupHC & 0.76 (0.61) & 0.79 (0.58) & 0.93 (0.64) & \textbf{0.94} (0.64) & 0.93 (0.64) & 0.93 (0.64) & \textbf{0.94} (0.64) & \textbf{0.94} (0.64) \\
Zscale & 0.78 (0.61) & 0.81 (0.59) & 0.92 (0.62) & 0.93 (0.63) & 0.82 (0.65) & 0.91 (0.64) & \textbf{0.94} (0.65)  & \textbf{0.94} (0.65)\\
\hline
\end{tabular}
\centering
\caption{Median bounding box IOUs for matched objects in the COSMOS HST test. The best bounding box IOU for each row is emphasized in bold.  Also shown in parentheses are the median segmentation mask IOUs. An IOU above 0.5 is considered to be a good match, and a score of 1.0 is a perfect overlap of ground truth and inference.}
\label{tab:resnet_ious}
\end{table*}

\begin{figure*}
	\includegraphics[width=\textwidth]{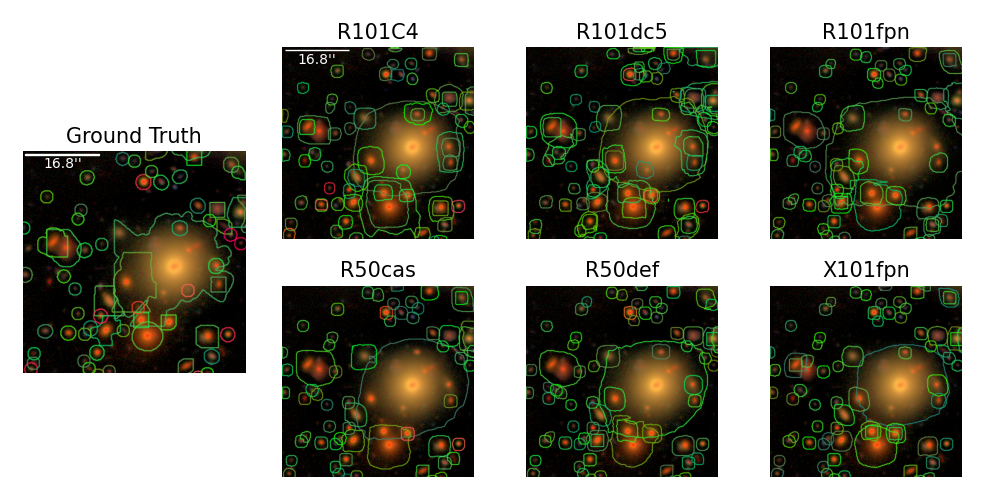}
    \caption{Inference on a close blend. The ground truth is shown on the left. RGB images are created with a Lupton contrast scaling. Other panels show model inference of segmentation maps and classes.  Top row, left to right: R101C4, R101dc5, R101fpn. Bottom row, left to right: R50cas, R50def, X101fpn. The colors indicate classes, green for galaxy and red for star. Differences in detections are solely due to the different backbones.  While the networks do not pick up every ground truth object, they are also able to detect real objects that were missed by our ground truth labelling.}
    \label{fig:res_blend_1}
\end{figure*}

\begin{figure*}
	\includegraphics[width=\textwidth]{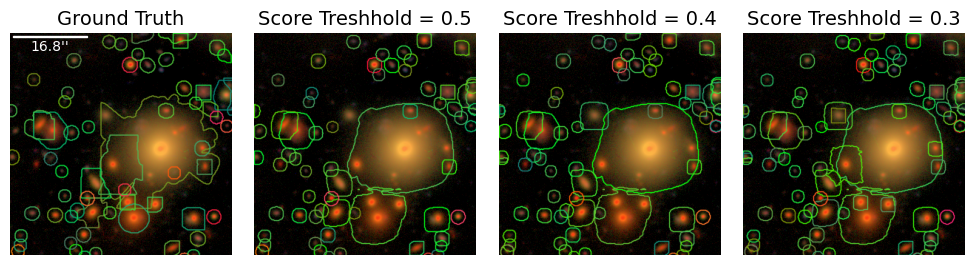}
    \caption{Inference on the same close blend as Figure \ref{fig:res_blend_1}, but only with a Swin architecture. The ground truth is shown on the left most panel, and the effect of lowering the detection confidence threshold to 0.5, 0.4, 0.3 is shown in left to right, respectively. As the threshold is lowered, objects within a larger footprint are detected.}
    \label{fig:swin_hyperparam_blend}
\end{figure*}

\begin{figure}
 	\includegraphics[width=\columnwidth]{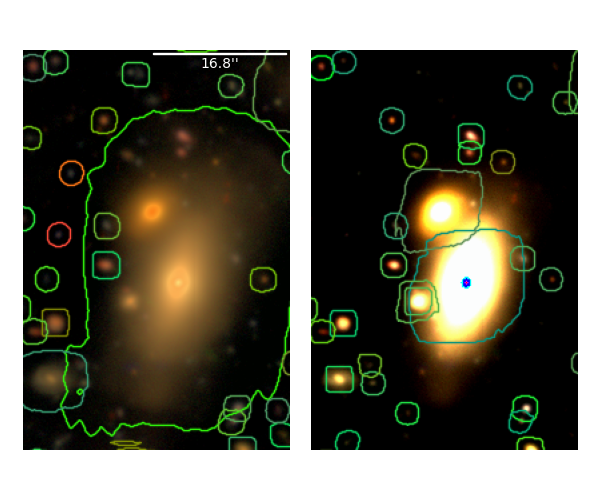}
    \caption{The effect of using a different contrast scaling on a close blend.  We show inference of a R50cas network when trained on Lupton scaled images (left) and z-scaled images (right).  The objects are more easily distinguished with a z-scaling.}
    \label{fig:R50cas_blend_con}
    
\end{figure}

The median IOUs measure the ability of the network to detect and segment objects, but it does not fully capture the deblending power of the networks. We examine the cases of a few close blends to get a sense of the ability of the networks to distinguish large overlapping objects.  We demonstrate the deblending capabilities of the different networks in Figure \ref{fig:res_blend_1}.  In very crowded scenes, the networks are able to distinguish the individual sources, and even pick up objects that are not present in the labelled set, which may present an advantage for studies of low surface-brightness galaxies.  As discussed in Section \ref{subsec:missing_label_bias}, this is likely due to the difference in object detection abilities of the Region Proposal Networks compared to peak-finding methods, and highlights that the models are not limited by the training data, but are able to extrapolate beyond it. It is also possible to alter inference hyperparameters such as IOU or detection confidence thresholds, which could allow for more or less detections or overlap between detections.  In Figure \ref{fig:swin_hyperparam_blend} we demonstrate the effect of lowering the  confidence threshold hyperparameter, allowing for more low-confidence detections.  While not equivalent, this is similar to lowering the detection threshold in peak-finding algorithms.  There are cases in which deblending is poor, and these are typically very large galaxies with one or more very large and very close companions. In such instances, it may be better to use a different contrast scaling. In Figure \ref{fig:R50cas_blend_con}, a Lupton contrast scaling prevents the network from deblending multiple large sources. With the same IOU/confidence score thresholds, a z-scaling works to better isolate the two sources. This is likely due to much larger dynamic range of our z-scaling, which allows for less smearing of the sources and more distinguishing power in this case.  Overall, there does not seem to be a one-size-fits-all network configuration for the cases of very large and very close blends.  Training on more data would likely improve the ability to detect and segment these objects. 

\section{Discussion}
\label{sec:discussion}
The effectiveness of instance segmentation models has been proven in many domains, boosted by the ability of networks to work ``out-of-the-box'' and without much fine-tuning. It has been shown that an object detection model based on the Mask R-CNN framework performs well in the classification and detection/segmentation of simulated astronomical survey images \citep{burke_deblending_2019}. In this work, we have trained and tested a broad range of state-of-the-art instance segmentation models on real data taken from the HSC SSP Data Release 3 to push the direction of deep learning based galaxy detection, classification, and deblending towards real applications. Network training and evaluation performance is limited by the efficacy of our label generation methodology, a task not easily formulated when the ground truth is not completely known. This limitation also affects the choices of metrics we use to measure network performance.  Often, classification and detection power are combined into the AP score, used throughout instance segmentation literature. However, this may not the best choice of metric for comparisons, as it implicitly assumes the completeness and correctness of the ground truth labels. To attempt to mitigate the effects of incorrect labels on performance metrics, we construct a test set of objects with class labels determined from more accurate space-based HST observations. However, since the AP metric artificially suffers from the detections of ``false positives'' that are true objects simply missing from the labelled set and/or the presence of spurious ground truth detections, we further attempt to mitigate this bias by constraining performance metrics to detected objects that have a matched ground truth label.

We find that all networks perform well at classifying galaxies, even out to the faintest in the sample.  Despite the wide variety of colors, sizes, and morphologies in the real imaging data, our models can identify these objects.  Stellar classification is worse, likely due to the smaller sample size in the training and test set.  Transformer based networks generally outperform ResNet based networks in classification power of both stars and galaxies.  They also appear to be more robust classifiers as magnitudes become fainter. Transformer based models maintain near 100\% completeness (recall) and purity (precision) of galaxy selection across the whole sample and above 60\% completeness and 80\% purity of stars out to i-band magnitudes of 25 mag.  These models are able to outperform the extendedness classifier used in the HSC catalogs, which depending on cuts yields near 100\% galaxy purity, roughly 90\% galaxy completeness, stellar completeness slightly above 50\% and stellar purity slightly above 40\% at i-band magnitudes of 25 mag \citep{bosch_hyper_2018}.  The performance increase of our models is especially noteworthy because they are able to surpass the HSC class labelling despite being trained with it. Transformer models are also more robust to different contrast scalings than traditional convolutional neural networks, indicating that they may be less susceptible to biases introduced from transfer learning with terrestrial images and more applicable to a wide range of images across surveys with different dynamic ranges.

The detection/deblending capabilities are measured by the median bounding box IOUs of the networks.  Again, transformer based networks generally outperform convolutional ResNet based networks.  The improved performance of transformer networks over convolutional based ones may be attributable to the ability of different attention heads to encode information at different image scale sizes \citep{Dosovitskiy20}, allowing for more overall global information propagation than CNNs.  While a convolutional neural network is able to learn spatial features through sliding a kernel across an image, a transformer learns features over the entire input at once, removing any limitations due to kernel sizes.  It is possible that the transformer backbones are implicitly utilizing large scale features in the images such as the spatial clustering of objects, background noise or seeing and using these bulk properties to inform the network.  

We examine a few cases of close blends to qualitatively see how the networks distinguish objects. There are cases in which the networks do not detect close objects, but these can sometimes be mitigated by altering the confidence and NMS IOU threshold hyperparmeters (which can be done after training).  In other cases, using a different contrast scaling helps to isolate closely blended objects. 

There is room to improve both classification and segmentation of these models in future work.  One possibility is constructing a larger training set with more accurate labels.  With better and larger samples of stars/galaxies, networks may perform better on classification. The more close blends of large galaxies seen during training, the more likely the networks will be able to distinguish these scenes.  There could be more fine-tuning of hyperparmeters done to the architectures before training, rather than running them out-of-the-box.  Additionally, the use of more photometric information could help in all tasks.  We use the \textit{i}, \textit{r} and \textit{g} bands on the HSC instrument in this work, corresponding to RGB color images, but could further investigate the performance if we include the \textit{z} and \textit{y} bands.  

It is possible that these networks need to be trained longer, or that the fundamentally different properties of astronomical images over terrestrial ones limits the abilities of these architectures in extracting useful features for classification. Despite our attempts to mitigate measurement biases arising from label generation, classification remains a challenge for these models at faint magnitudes.  A machine learning model has already been used to classify HSC data using photometry information with better accuracy than morphological methods, but relies on the upstream task of detection \citep{bosch_hyper_2018}. The instance segmentation models presented in this work are able to identify and assign classes after training using only an image as input.  

\section{Conclusions}
\label{sec:conclusions}

It is already a necessary consequence of the current epoch of astronomical research for machine learning algorithms to parse through massive sets of images. A first step in catalog construction is detecting these objects from imaging data.  Advancements in the broader computer vision community have given rise to a large ecosystem of models that perform many necessary tasks at once, including detection, segmentation, and classification.  While tried and tested on terrestrial data and shown to work on simulated astronomical data, the application on real survey images remains a work in progress.  Many methods rely on the object detection stage to produce measurements of individual objects. In this work, we employ a variety of instance segmentation models available through \textsc{Detectron2} to perform the detection task as well as deblending and object classification simultaneously on images taken from the HSC-SSP Data Release 3.  We carefully construct ground truth labels with existing frameworks and catalog matching, and caution that real data gives no straightforward way of producing labels. We find that the best networks perform well at classifying the faintest galaxies in the sample, and perform better than traditional methods at classifying stars up to \textit{i}-band magnitudes of $\sim$25 mag.  We find that even if trained on less accurate class labels, the neural networks still pick up on useful features that allow inference of the true underlying class.  We expect more data with accurate labels to improve performance.  The best performing models are able to detect and deblend by matching ground truth object locations and bounding boxes. Transformer networks appear to be a promising avenue of exploration in further studies.  

There are many other areas for future study. While we tested a variety of models, there are many within \textsc{Detectron2} that we did not implement. Some architectures are quite large and require significant resources to train. For example, we attempted to implement ViT backbones \citep{Dosovitskiy20} among our set of transformer-based architectures, but were limited by the available GPU memory.  Many models, especially transformers, are trained with state-of-the-art computing resources at FAIR or other organizations, and subsequently retraining them demands significant resources.  Tests could be done on other sets of real data, with other downstream tasks in mind.  For example, \cite{Gonzalez2018} investigate the application of instance segmentation models on SDSS data to classify galaxy morphologies.  It would be straightforward to add additional classes, or implement a redshift estimation network using the modular nature of \textsc{Detectron2}.  In future work we plan to add a photo-z estimator branch to the Mask R-CNN/transformer networks and interface with the LSST software RAIL (Redshift Assessment Infrastructure Layers)\footnote{\hyperlink{https://github.com/LSSTDESC/RAIL}{https://github.com/LSSTDESC/RAIL}}.  The availability of realistic LSST-like simulations \citep{DC2} for training will allow us to avoid biases from label generation. The efficiency of neural networks and the ability to perform multiple tasks at once is now a necessity with the amount of survey data pouring into pipelines.

As surveys push deeper into the sky, they will produce unprecedented amounts of objects that will be necessary to process.  LSST will provide the deepest ground-based observations ever, and survey terrabytes of data every night, highlighting a need for accurate and precise object detection and classification, potentially in real-time. Correctly classifying and and deblending sources will be necessary for a wide range of studies, and deep instance segmentation models will be a valuable tool in handling these tasks.

\section*{Acknowledgements}

We thank Dr. S. Luo and Dr. D. Mu at the National Center for Supercomputing Applications (NCSA) for their assistance with the GPU cluster used in this work. We thank Y. Shen for helpful discussion on the HST observations of the COSMOS field.
%We thank the anonymous referees for helpful comments. \pat{We didn't submit this yet, so we can't thank the referee... :P} 
G.M., Y.L., Y.L. and X.L. acknowledge support from the NCSA Faculty Fellowship and the NCSA SPIN programs. %and NSF grant AST-2308174.  

This work utilizes resources supported by the National Science Foundation's Major Research Instrumentation program, grant \#1725729, as well as the University of Illinois at Urbana-Champaign. 

We acknowledge use of Matplotlib \citep{Hunter2007}, a community-developed Python library for plotting. This research made use of Astropy,\footnote{\hyperlink{http://www.astropy.org}{http://www.astropy.org}} a community-developed core Python package for Astronomy \citep{astropy:2013, astropy:2018}. This research has made use of NASA's Astrophysics Data System.

The Hyper Suprime-Cam (HSC) collaboration includes the astronomical communities of Japan and Taiwan, and Princeton University. The HSC instrumentation and software were developed by the National Astronomical Observatory of Japan (NAOJ), the Kavli Institute for the Physics and Mathematics of the Universe (Kavli IPMU), the University of Tokyo, the High Energy Accelerator Research Organization (KEK), the Academia Sinica Institute for Astronomy and Astrophysics in Taiwan (ASIAA), and Princeton University. Funding was contributed by the FIRST program from Japanese Cabinet Office, the Ministry of Education, Culture, Sports, Science and Technology (MEXT), the Japan Society for the Promotion of Science (JSPS), Japan Science and Technology Agency (JST), the Toray Science Foundation, NAOJ, Kavli IPMU, KEK, ASIAA, and Princeton University. 

This paper makes use of software developed for the Large Synoptic Survey Telescope. We thank the LSST Project for making their code available as free software at  http://dm.lsst.org

The Pan-STARRS1 Surveys (PS1) have been made possible through contributions of the Institute for Astronomy, the University of Hawaii, the Pan-STARRS Project Office, the Max-Planck Society and its participating institutes, the Max Planck Institute for Astronomy, Heidelberg and the Max Planck Institute for Extraterrestrial Physics, Garching, The Johns Hopkins University, Durham University, the University of Edinburgh, Queen’s University Belfast, the Harvard-Smithsonian Center for Astrophysics, the Las Cumbres Observatory Global Telescope Network Incorporated, the National Central University of Taiwan, the Space Telescope Science Institute, the National Aeronautics and Space Administration under Grant No. NNX08AR22G issued through the Planetary Science Division of the NASA Science Mission Directorate, the National Science Foundation under Grant No. AST-1238877, the University of Maryland, and Eotvos Lorand University (ELTE) and the Los Alamos National Laboratory.

Based [in part] on data collected at the Subaru Telescope and retrieved from the HSC data archive system, which is operated by Subaru Telescope and Astronomy Data Center at National Astronomical Observatory of Japan.

This research has made use of the NASA/IPAC Infrared Science Archive, which is funded by the National Aeronautics and Space Administration and operated by the California Institute of Technology.

\section*{Data Availability}
The data underlying this article were accessed from the HSC data archive system https://hsc-release.mtk.nao.ac.jp/doc/. The derived data generated in this research will be shared on reasonable request to the corresponding author. The software used in this work is publicly available at https://github.com/grantmerz/deepdisc.

%%%%%%%%%%%%%%%%%%%%%%%%%%%%%%%%%%%%%%%%%%%%%%%%%%



\bibliographystyle{mnras}
\bibliography{ref} % if your bibtex file is called ref.bib

\begin{thebibliography}{}
\makeatletter
\relax
\def\mn@urlcharsother{\let\do\@makeother \do\$\do\&\do\#\do\^\do\_\do\%\do\~}
\def\mn@doi{\begingroup\mn@urlcharsother \@ifnextchar [ {\mn@doi@}
  {\mn@doi@[]}}
\def\mn@doi@[#1]#2{\def\@tempa{#1}\ifx\@tempa\@empty \href
  {http://dx.doi.org/#2} {doi:#2}\else \href {http://dx.doi.org/#2} {#1}\fi
  \endgroup}
\def\mn@eprint#1#2{\mn@eprint@#1:#2::\@nil}
\def\mn@eprint@arXiv#1{\href {http://arxiv.org/abs/#1} {{\tt arXiv:#1}}}
\def\mn@eprint@dblp#1{\href {http://dblp.uni-trier.de/rec/bibtex/#1.xml}
  {dblp:#1}}
\def\mn@eprint@#1:#2:#3:#4\@nil{\def\@tempa {#1}\def\@tempb {#2}\def\@tempc
  {#3}\ifx \@tempc \@empty \let \@tempc \@tempb \let \@tempb \@tempa \fi \ifx
  \@tempb \@empty \def\@tempb {arXiv}\fi \@ifundefined
  {mn@eprint@\@tempb}{\@tempb:\@tempc}{\expandafter \expandafter \csname
  mn@eprint@\@tempb\endcsname \expandafter{\@tempc}}}

\bibitem[\protect\citeauthoryear{Aihara et~al.,}{Aihara
  et~al.}{2018a}]{aihara_hyper_2018}
Aihara H.,  et~al., 2018a, \mn@doi [Publications of the Astronomical Society of
  Japan] {10.1093/pasj/psx066}, 70, S4

\bibitem[\protect\citeauthoryear{{Aihara} et~al.,}{{Aihara}
  et~al.}{2018b}]{Aihara2018}
{Aihara} H.,  et~al., 2018b, \mn@doi [\pasj] {10.1093/pasj/psx081}, \href
  {https://ui.adsabs.harvard.edu/abs/2018PASJ...70S...8A} {70, S8}

\bibitem[\protect\citeauthoryear{Aihara et~al.,}{Aihara
  et~al.}{2022}]{aihara_third_2022}
Aihara H.,  et~al., 2022, \mn@doi [Publications of the Astronomical Society of
  Japan] {10.1093/pasj/psab122}, 74, 247

\bibitem[\protect\citeauthoryear{{Alam} et~al.,}{{Alam} et~al.}{2015}]{SDSS}
{Alam} S.,  et~al., 2015, \mn@doi [\apjs] {10.1088/0067-0049/219/1/12}, \href
  {https://ui.adsabs.harvard.edu/abs/2015ApJS..219...12A} {219, 12}

\bibitem[\protect\citeauthoryear{{Amiaux} et~al.}{{Amiaux}
  et~al.}{2012}]{Amiaux12}
{Amiaux} J.,  et~al., 2012, Euclid Mission: building of a reference survey,
  \mn@doi{10.1117/12.926513}, \url {https://doi.org/10.1117/12.926513}

\bibitem[\protect\citeauthoryear{{Andreon}, {Gargiulo}, {Longo}, {Tagliaferri}
  \& {Capuano}}{{Andreon} et~al.}{2000}]{Andreon00}
{Andreon} S.,  {Gargiulo} G.,  {Longo} G.,  {Tagliaferri} R.,   {Capuano} N.,
  2000, \mn@doi [\mnras] {10.1046/j.1365-8711.2000.03700.x}, \href
  {https://ui.adsabs.harvard.edu/abs/2000MNRAS.319..700A} {319, 700}

\bibitem[\protect\citeauthoryear{Arcelin, Doux, Aubourg, Roucelle  \& {LSST
  Dark Energy Science Collaboration}}{Arcelin
  et~al.}{2021}]{arcelin_deblending_2021}
Arcelin B.,  Doux C.,  Aubourg E.,  Roucelle C.,   {LSST Dark Energy Science
  Collaboration} 2021, \mn@doi [Monthly Notices of the Royal Astronomical
  Society] {10.1093/mnras/staa3062}, 500, 531

\bibitem[\protect\citeauthoryear{{Astropy Collaboration} et~al.,}{{Astropy
  Collaboration} et~al.}{2013}]{astropy:2013}
{Astropy Collaboration} et~al., 2013, \mn@doi [\aap]
  {10.1051/0004-6361/201322068}, \href
  {http://adsabs.harvard.edu/abs/2013A%26A...558A..33A} {558, A33}

\bibitem[\protect\citeauthoryear{{Bertin} \& {Arnouts}}{{Bertin} \&
  {Arnouts}}{1996}]{SExtractor}
{Bertin} E.,  {Arnouts} S.,  1996, \mn@doi [\aaps] {10.1051/aas:1996164}, \href
  {https://ui.adsabs.harvard.edu/abs/1996A&AS..117..393B} {117, 393}

\bibitem[\protect\citeauthoryear{{Bochkovskiy}, {Wang}  \&
  {Liao}}{{Bochkovskiy} et~al.}{2020}]{YOLOv4}
{Bochkovskiy} A.,  {Wang} C.-Y.,   {Liao} H.-Y.~M.,  2020, \mn@doi [arXiv
  e-prints] {10.48550/arXiv.2004.10934}, \href
  {https://ui.adsabs.harvard.edu/abs/2020arXiv200410934B} {p. arXiv:2004.10934}

\bibitem[\protect\citeauthoryear{Bosch et~al.,}{Bosch
  et~al.}{2018}]{bosch_hyper_2018}
Bosch J.,  et~al., 2018, \mn@doi [Publications of the Astronomical Society of
  Japan] {10.1093/pasj/psx080}, 70, S5

\bibitem[\protect\citeauthoryear{{Boucaud} et~al.,}{{Boucaud}
  et~al.}{2020}]{Boucaud19}
{Boucaud} A.,  et~al., 2020, \mn@doi [\mnras] {10.1093/mnras/stz3056}, \href
  {https://ui.adsabs.harvard.edu/abs/2020MNRAS.491.2481B} {491, 2481}

\bibitem[\protect\citeauthoryear{Bretonnière, Boucaud  \&
  Huertas-Company}{Bretonnière et~al.}{2021}]{bretonniere_probabilistic_2021}
Bretonnière H.,  Boucaud A.,   Huertas-Company M.,  2021, Probabilistic
  segmentation of overlapping galaxies for large cosmological surveys,
  \mn@doi{10.48550/arXiv.2111.15455}, \url {http://arxiv.org/abs/2111.15455}

\bibitem[\protect\citeauthoryear{{Burke}, {Aleo}, {Chen}, {Liu}, {Peterson},
  {Sembroski}  \& {Lin}}{{Burke} et~al.}{2019}]{burke_deblending_2019}
{Burke} C.~J.,  {Aleo} P.~D.,  {Chen} Y.-C.,  {Liu} X.,  {Peterson} J.~R.,
  {Sembroski} G.~H.,   {Lin} J. Y.-Y.,  2019, \mn@doi [\mnras]
  {10.1093/mnras/stz2845}, \href
  {https://ui.adsabs.harvard.edu/abs/2019MNRAS.490.3952B} {490, 3952}

\bibitem[\protect\citeauthoryear{Cai \& Vasconcelos}{Cai \&
  Vasconcelos}{2018}]{Cai17}
Cai Z.,  Vasconcelos N.,  2018, in Proceedings of the IEEE conference on
  computer vision and pattern recognition. pp 6154--6162

\bibitem[\protect\citeauthoryear{Caron, Touvron, Misra, J{\'e}gou, Mairal,
  Bojanowski  \& Joulin}{Caron et~al.}{2021}]{DINOv1}
Caron M.,  Touvron H.,  Misra I.,  J{\'e}gou H.,  Mairal J.,  Bojanowski P.,
  Joulin A.,  2021, in Proceedings of the IEEE/CVF international conference on
  computer vision. pp 9650--9660

\bibitem[\protect\citeauthoryear{Cheng}{Cheng}{2017}]{Cheng2017}
Cheng J.,  2017, PhD thesis, Purdue University

\bibitem[\protect\citeauthoryear{Cheng, Parkhi  \& Kirillov}{Cheng
  et~al.}{2022}]{maskformer21}
Cheng B.,  Parkhi O.,   Kirillov A.,  2022, in Proceedings of the IEEE/CVF
  Conference on Computer Vision and Pattern Recognition. pp 2617--2626

\bibitem[\protect\citeauthoryear{{Dai}, {Qi}, {Xiong}, {Li}, {Zhang}, {Hu}  \&
  {Wei}}{{Dai} et~al.}{2017}]{Dai17}
{Dai} J.,  {Qi} H.,  {Xiong} Y.,  {Li} Y.,  {Zhang} G.,  {Hu} H.,   {Wei} Y.,
  2017, \mn@doi [arXiv e-prints] {10.48550/arXiv.1703.06211}, \href
  {https://ui.adsabs.harvard.edu/abs/2017arXiv170306211D} {p. arXiv:1703.06211}

\bibitem[\protect\citeauthoryear{{Dark Energy Survey Collaboration}
  et~al.,}{{Dark Energy Survey Collaboration} et~al.}{2016}]{DES16}
{Dark Energy Survey Collaboration} et~al., 2016, \mn@doi [\mnras]
  {10.1093/mnras/stw641}, \href
  {https://ui.adsabs.harvard.edu/abs/2016MNRAS.460.1270D} {460, 1270}

\bibitem[\protect\citeauthoryear{{Dawson}, {Schneider}, {Tyson}  \&
  {Jee}}{{Dawson} et~al.}{2016}]{Dawson16}
{Dawson} W.~A.,  {Schneider} M.~D.,  {Tyson} J.~A.,   {Jee} M.~J.,  2016,
  \mn@doi [\apj] {10.3847/0004-637X/816/1/11}, \href
  {http://adsabs.harvard.edu/abs/2016ApJ...816...11D} {816, 11}

\bibitem[\protect\citeauthoryear{Deng, Dong, Socher, Li, Li  \& Fei-Fei}{Deng
  et~al.}{2009}]{ImageNet}
Deng J.,  Dong W.,  Socher R.,  Li L.-J.,  Li K.,   Fei-Fei L.,  2009, in 2009
  IEEE Conference on Computer Vision and Pattern Recognition. pp 248--255,
  \mn@doi{10.1109/CVPR.2009.5206848}

\bibitem[\protect\citeauthoryear{{Dey} et~al.,}{{Dey} et~al.}{2019}]{Dey2019}
{Dey} A.,  et~al., 2019, \mn@doi [\aj] {10.3847/1538-3881/ab089d}, \href
  {https://ui.adsabs.harvard.edu/abs/2019AJ....157..168D} {157, 168}

\bibitem[\protect\citeauthoryear{{Dosovitskiy} et~al.,}{{Dosovitskiy}
  et~al.}{2020}]{Dosovitskiy20}
{Dosovitskiy} A.,  et~al., 2020, \mn@doi [arXiv e-prints]
  {10.48550/arXiv.2010.11929}, \href
  {https://ui.adsabs.harvard.edu/abs/2020arXiv201011929D} {p. arXiv:2010.11929}

\bibitem[\protect\citeauthoryear{Fan, Xiong, Mangalam, Li, Yan, Malik  \&
  Feichtenhofer}{Fan et~al.}{2021}]{Fan21}
Fan H.,  Xiong B.,  Mangalam K.,  Li Y.,  Yan Z.,  Malik J.,   Feichtenhofer
  C.,  2021, in Proceedings of the IEEE/CVF international conference on
  computer vision. pp 6824--6835

\bibitem[\protect\citeauthoryear{{Flaugher} et~al.,}{{Flaugher}
  et~al.}{2015}]{DECam}
{Flaugher} B.,  et~al., 2015, \mn@doi [\aj] {10.1088/0004-6256/150/5/150},
  \href {https://ui.adsabs.harvard.edu/abs/2015AJ....150..150F} {150, 150}

\bibitem[\protect\citeauthoryear{{Girshick}}{{Girshick}}{2015}]{Girshick2015}
{Girshick} R.,  2015, in 2015 IEEE International Conference on Computer Vision
  (ICCV). pp 1440--1448, \mn@doi{10.1109/ICCV.2015.169}

\bibitem[\protect\citeauthoryear{{Gonz{\'a}lez}, {Mu{\~n}oz}  \&
  {Hern{\'a}ndez}}{{Gonz{\'a}lez} et~al.}{2018}]{Gonzalez2018}
{Gonz{\'a}lez} R.~E.,  {Mu{\~n}oz} R.~P.,   {Hern{\'a}ndez} C.~A.,  2018,
  \mn@doi [Astronomy and Computing] {10.1016/j.ascom.2018.09.004}, \href
  {https://ui.adsabs.harvard.edu/abs/2018A&C....25..103G} {25, 103}

\bibitem[\protect\citeauthoryear{{Grogin} et~al.,}{{Grogin}
  et~al.}{2011}]{CANDELS1}
{Grogin} N.~A.,  et~al., 2011, \mn@doi [\apjs] {10.1088/0067-0049/197/2/35},
  \href {https://ui.adsabs.harvard.edu/abs/2011ApJS..197...35G} {197, 35}

\bibitem[\protect\citeauthoryear{Hausen \& Robertson}{Hausen \&
  Robertson}{2020}]{hausen_morpheus_2020}
Hausen R.,  Robertson B.,  2020, \mn@doi [ApJS] {10.3847/1538-4365/ab8868},
  248, 20

\bibitem[\protect\citeauthoryear{He, Zhang, Ren  \& Sun}{He
  et~al.}{2016}]{He2016}
He K.,  Zhang X.,  Ren S.,   Sun J.,  2016, 2016 IEEE Conference on Computer
  Vision and Pattern Recognition (CVPR), pp 770--778

\bibitem[\protect\citeauthoryear{He, Gkioxari, Doll{\'a}r  \& Girshick}{He
  et~al.}{2017}]{he_mask_2018}
He K.,  Gkioxari G.,  Doll{\'a}r P.,   Girshick R.,  2017, in Proceedings of
  the IEEE international conference on computer vision. pp 2961--2969

\bibitem[\protect\citeauthoryear{He, Qiu, Luo, Shi, Kong  \& Jiang}{He
  et~al.}{2021}]{He21}
He Z.,  Qiu B.,  Luo A.-L.,  Shi J.,  Kong X.,   Jiang X.,  2021, \mn@doi
  [Monthly Notices of the Royal Astronomical Society] {10.1093/mnras/stab2243},
  508, 2039

\bibitem[\protect\citeauthoryear{{Hemmati} et~al.,}{{Hemmati}
  et~al.}{2022}]{Hemmati22}
{Hemmati} S.,  et~al., 2022, \mn@doi [\apj] {10.3847/1538-4357/aca1b8}, \href
  {https://ui.adsabs.harvard.edu/abs/2022ApJ...941..141H} {941, 141}

\bibitem[\protect\citeauthoryear{{Huertas-Company} \&
  {Lanusse}}{{Huertas-Company} \& {Lanusse}}{2023}]{Dawes23}
{Huertas-Company} M.,  {Lanusse} F.,  2023, \mn@doi [\pasa]
  {10.1017/pasa.2022.55}, \href
  {https://ui.adsabs.harvard.edu/abs/2023PASA...40....1H} {40, e001}

\bibitem[\protect\citeauthoryear{Hunter}{Hunter}{2007}]{Hunter2007}
Hunter J.~D.,  2007, \mn@doi [Computing in Science \& Engineering]
  {10.1109/MCSE.2007.55}, 9, 90

\bibitem[\protect\citeauthoryear{Ibrahim, Haworth  \& Cheng}{Ibrahim
  et~al.}{2020}]{IBRAHIM20}
Ibrahim M.~R.,  Haworth J.,   Cheng T.,  2020, \mn@doi [Cities]
  {https://doi.org/10.1016/j.cities.2019.102481}, 96, 102481

\bibitem[\protect\citeauthoryear{{Ivezi{\'c}} et~al.,}{{Ivezi{\'c}}
  et~al.}{2019}]{Ivezic2019}
{Ivezi{\'c}} {\v{Z}}.,  et~al., 2019, \mn@doi [\apj]
  {10.3847/1538-4357/ab042c}, \href
  {https://ui.adsabs.harvard.edu/abs/2019ApJ...873..111I} {873, 111}

\bibitem[\protect\citeauthoryear{{Jarvis} \& {Tyson}}{{Jarvis} \&
  {Tyson}}{1981}]{Jarvis81}
{Jarvis} J.~F.,  {Tyson} J.~A.,  1981, \mn@doi [\aj] {10.1086/112907}, \href
  {https://ui.adsabs.harvard.edu/abs/1981AJ.....86..476J} {86, 476}

\bibitem[\protect\citeauthoryear{Kawanomoto et~al.,}{Kawanomoto
  et~al.}{2018}]{Kawanomoto2018}
Kawanomoto S.,  et~al., 2018, \mn@doi [\pasj] {10.1093/pasj/psy056}, 70, 66

\bibitem[\protect\citeauthoryear{Kindratenko et~al.,}{Kindratenko
  et~al.}{2020}]{10.1145/3311790.3396649}
Kindratenko V.,  et~al., 2020, in Practice and Experience in Advanced Research
  Computing. PEARC '20.
Association for Computing Machinery, New York, NY, USA, p. 41–48,
  \mn@doi{10.1145/3311790.3396649}

\bibitem[\protect\citeauthoryear{{Koekemoer} et~al.,}{{Koekemoer}
  et~al.}{2011}]{CANDELS2}
{Koekemoer} A.~M.,  et~al., 2011, \mn@doi [\apjs] {10.1088/0067-0049/197/2/36},
  \href {https://ui.adsabs.harvard.edu/abs/2011ApJS..197...36K} {197, 36}

\bibitem[\protect\citeauthoryear{{Kroupa}}{{Kroupa}}{2001}]{Kroupa2001}
{Kroupa} P.,  2001, \mn@doi [\mnras] {10.1046/j.1365-8711.2001.04022.x}, \href
  {https://ui.adsabs.harvard.edu/abs/2001MNRAS.322..231K} {322, 231}

\bibitem[\protect\citeauthoryear{{LSST Dark Energy Science Collaboration (LSST
  DESC)} et~al.,}{{LSST Dark Energy Science Collaboration (LSST DESC)}
  et~al.}{2021}]{DC2}
{LSST Dark Energy Science Collaboration (LSST DESC)} et~al., 2021, \mn@doi
  [\apjs] {10.3847/1538-4365/abd62c}, \href
  {https://ui.adsabs.harvard.edu/abs/2021ApJS..253...31L} {253, 31}

\bibitem[\protect\citeauthoryear{{Leauthaud} et~al.,}{{Leauthaud}
  et~al.}{2007}]{Leauthaud07}
{Leauthaud} A.,  et~al., 2007, \mn@doi [\apjs] {10.1086/516598}, \href
  {https://ui.adsabs.harvard.edu/abs/2007ApJS..172..219L} {172, 219}

\bibitem[\protect\citeauthoryear{Li, Wu, Fan, Mangalam, Xiong, Malik  \&
  Feichtenhofer}{Li et~al.}{2022}]{Li21}
Li Y.,  Wu C.-Y.,  Fan H.,  Mangalam K.,  Xiong B.,  Malik J.,   Feichtenhofer
  C.,  2022, in Proceedings of the IEEE/CVF Conference on Computer Vision and
  Pattern Recognition. pp 4804--4814

\bibitem[\protect\citeauthoryear{Lin, Maire, Belongie, Hays, Perona, Ramanan,
  Doll{\'a}r  \& Zitnick}{Lin et~al.}{2014}]{Lin2014}
Lin T.-Y.,  Maire M.,  Belongie S.,  Hays J.,  Perona P.,  Ramanan D.,
  Doll{\'a}r P.,   Zitnick C.~L.,  2014, in European Conference on Computer
  Vision (ECCV). Z{\"u}rich

\bibitem[\protect\citeauthoryear{Lin, Doll{\'a}r, Girshick, He, Hariharan  \&
  Belongie}{Lin et~al.}{2017}]{Lin17}
Lin T.-Y.,  Doll{\'a}r P.,  Girshick R.~B.,  He K.,  Hariharan B.,   Belongie
  S.~J.,  2017, 2017 IEEE Conference on Computer Vision and Pattern Recognition
  (CVPR), pp 936--944

\bibitem[\protect\citeauthoryear{{Lintott} et~al.,}{{Lintott}
  et~al.}{2011}]{Lintott11}
{Lintott} C.,  et~al., 2011, \mn@doi [\mnras]
  {10.1111/j.1365-2966.2010.17432.x}, \href
  {https://ui.adsabs.harvard.edu/abs/2011MNRAS.410..166L} {410, 166}

\bibitem[\protect\citeauthoryear{Liu, Lin, Cao, Hu, Wei, Zhang, Lin  \&
  Guo}{Liu et~al.}{2021}]{Liu21}
Liu Z.,  Lin Y.,  Cao Y.,  Hu H.,  Wei Y.,  Zhang Z.,  Lin S.,   Guo B.,  2021,
  in Proceedings of the IEEE/CVF international conference on computer vision.
  pp 10012--10022

\bibitem[\protect\citeauthoryear{{Lupton}, {Blanton}, {Fekete}, {Hogg},
  {O'Mullane}, {Szalay}  \& {Wherry}}{{Lupton} et~al.}{2004}]{Lupton2004}
{Lupton} R.,  {Blanton} M.~R.,  {Fekete} G.,  {Hogg} D.~W.,  {O'Mullane} W.,
  {Szalay} A.,   {Wherry} N.,  2004, \mn@doi [\pasp] {10.1086/382245}, \href
  {https://ui.adsabs.harvard.edu/abs/2004PASP..116..133L} {116, 133}

\bibitem[\protect\citeauthoryear{{Madau} \& {Dickinson}}{{Madau} \&
  {Dickinson}}{2014}]{Madau2014}
{Madau} P.,  {Dickinson} M.,  2014, \mn@doi [\araa]
  {10.1146/annurev-astro-081811-125615}, \href
  {https://ui.adsabs.harvard.edu/abs/2014ARA&A..52..415M} {52, 415}

\bibitem[\protect\citeauthoryear{{Mahabal} et~al.,}{{Mahabal}
  et~al.}{2019}]{Mahabal19}
{Mahabal} A.,  et~al., 2019, \mn@doi [\pasp] {10.1088/1538-3873/aaf3fa}, \href
  {https://ui.adsabs.harvard.edu/abs/2019PASP..131c8002M} {131, 038002}

\bibitem[\protect\citeauthoryear{{Malanchev} et~al.,}{{Malanchev}
  et~al.}{2021}]{Malanchev21}
{Malanchev} K.~L.,  et~al., 2021, \mn@doi [\mnras] {10.1093/mnras/stab316},
  \href {https://ui.adsabs.harvard.edu/abs/2021MNRAS.502.5147M} {502, 5147}

\bibitem[\protect\citeauthoryear{Melchior, Moolekamp, Jerdee, Armstrong, Sun,
  Bosch  \& Lupton}{Melchior et~al.}{2018}]{melchior_scarlet_2018}
Melchior P.,  Moolekamp F.,  Jerdee M.,  Armstrong R.,  Sun A.-L.,  Bosch J.,
  Lupton R.,  2018, \mn@doi [Astronomy and Computing]
  {10.1016/j.ascom.2018.07.001}, 24, 129

\bibitem[\protect\citeauthoryear{Melchior, Joseph, Sanchez, MacCrann  \&
  Gruen}{Melchior et~al.}{2021}]{melchior_challenge_2021}
Melchior P.,  Joseph R.,  Sanchez J.,  MacCrann N.,   Gruen D.,  2021, \mn@doi
  [Nat Rev Phys] {10.1038/s42254-021-00353-y}, 3, 712

\bibitem[\protect\citeauthoryear{{Miller} \& {Hall}}{{Miller} \&
  {Hall}}{2021}]{Miller21}
{Miller} A.~A.,  {Hall} X.~J.,  2021, \mn@doi [\pasp]
  {10.1088/1538-3873/abf038}, \href
  {https://ui.adsabs.harvard.edu/abs/2021PASP..133e4502M} {133, 054502}

\bibitem[\protect\citeauthoryear{Miyazaki et~al.,}{Miyazaki
  et~al.}{2017}]{Miyazaki18}
Miyazaki S.,  et~al., 2017, \mn@doi [Publications of the Astronomical Society
  of Japan] {10.1093/pasj/psx063}, 70

\bibitem[\protect\citeauthoryear{{Morganson} et~al.,}{{Morganson}
  et~al.}{2018}]{Morganson2018}
{Morganson} E.,  et~al., 2018, \mn@doi [\pasp] {10.1088/1538-3873/aab4ef},
  \href {https://ui.adsabs.harvard.edu/abs/2018PASP..130g4501M} {130, 074501}

\bibitem[\protect\citeauthoryear{{Muyskens}, {Goumiri}, {Priest}, {Schneider},
  {Armstrong}, {Bernstein}  \& {Dana}}{{Muyskens} et~al.}{2022}]{Muyskens22}
{Muyskens} A.~L.,  {Goumiri} I.~R.,  {Priest} B.~W.,  {Schneider} M.~D.,
  {Armstrong} R.~E.,  {Bernstein} J.,   {Dana} R.,  2022, \mn@doi [\aj]
  {10.3847/1538-3881/ac4e93}, \href
  {https://ui.adsabs.harvard.edu/abs/2022AJ....163..148M} {163, 148}

\bibitem[\protect\citeauthoryear{{Oquab} et~al.,}{{Oquab}
  et~al.}{2023}]{DINOv2}
{Oquab} M.,  et~al., 2023, \mn@doi [arXiv e-prints]
  {10.48550/arXiv.2304.07193}, \href
  {https://ui.adsabs.harvard.edu/abs/2023arXiv230407193O} {p. arXiv:2304.07193}

\bibitem[\protect\citeauthoryear{Pavel, Tan  \& Abdullah}{Pavel
  et~al.}{2022}]{Pavel22}
Pavel M.~I.,  Tan S.~Y.,   Abdullah A.,  2022, \mn@doi [Applied Sciences]
  {10.3390/app12146831}, 12

\bibitem[\protect\citeauthoryear{{Peterson} et~al.,}{{Peterson}
  et~al.}{2015}]{Peterson2015}
{Peterson} J.~R.,  et~al., 2015, \mn@doi [\apjs] {10.1088/0067-0049/218/1/14},
  \href {https://ui.adsabs.harvard.edu/abs/2015ApJS..218...14P} {218, 14}

\bibitem[\protect\citeauthoryear{{Price-Whelan} et~al.,}{{Price-Whelan}
  et~al.}{2018}]{astropy:2018}
{Price-Whelan} A.~M.,  et~al., 2018, \mn@doi [\aj] {10.3847/1538-3881/aabc4f},
  \href {https://ui.adsabs.harvard.edu/#abs/2018AJ....156..123T} {156, 123}

\bibitem[\protect\citeauthoryear{Reiman \& Göhre}{Reiman \&
  Göhre}{2019}]{reiman_deblending_2019}
Reiman D.~M.,  Göhre B.~E.,  2019, \mn@doi [Monthly Notices of the Royal
  Astronomical Society] {10.1093/mnras/stz575}, 485, 2617

\bibitem[\protect\citeauthoryear{{Ross} et~al.,}{{Ross} et~al.}{2011}]{Ross11}
{Ross} A.~J.,  et~al., 2011, \mn@doi [\mnras]
  {10.1111/j.1365-2966.2011.19351.x}, \href
  {https://ui.adsabs.harvard.edu/abs/2011MNRAS.417.1350R} {417, 1350}

\bibitem[\protect\citeauthoryear{{Russeil}, {Ishida}, {Le Montagner}, {Peloton}
   \& {Moller}}{{Russeil} et~al.}{2022}]{Russeil22}
{Russeil} E.,  {Ishida} E. E.~O.,  {Le Montagner} R.,  {Peloton} J.,   {Moller}
  A.,  2022, \mn@doi [arXiv e-prints] {10.48550/arXiv.2211.10987}, \href
  {https://ui.adsabs.harvard.edu/abs/2022arXiv221110987R} {p. arXiv:2211.10987}

\bibitem[\protect\citeauthoryear{{Scoville} et~al.,}{{Scoville}
  et~al.}{2007}]{COSMOS}
{Scoville} N.,  et~al., 2007, \mn@doi [\apjs] {10.1086/516585}, \href
  {https://ui.adsabs.harvard.edu/abs/2007ApJS..172....1S} {172, 1}

\bibitem[\protect\citeauthoryear{{Spergel} et~al.,}{{Spergel}
  et~al.}{2013}]{Spergel13}
{Spergel} D.,  et~al., 2013, arXiv e-prints, \href
  {https://ui.adsabs.harvard.edu/abs/2013arXiv1305.5422S} {p. arXiv:1305.5422}

\bibitem[\protect\citeauthoryear{{Tachibana} \& {Miller}}{{Tachibana} \&
  {Miller}}{2018}]{Tachibana18}
{Tachibana} Y.,  {Miller} A.~A.,  2018, \mn@doi [\pasp]
  {10.1088/1538-3873/aae3d9}, \href
  {https://ui.adsabs.harvard.edu/abs/2018PASP..130l8001T} {130, 128001}

\bibitem[\protect\citeauthoryear{Tan, Sun, Kong, Zhang, Yang  \& Liu}{Tan
  et~al.}{2018}]{Tan2018}
Tan C.,  Sun F.,  Kong T.,  Zhang W.,  Yang C.,   Liu C.,  2018, A Survey on
  Deep Transfer Learning: 27th International Conference on Artificial Neural
  Networks, Rhodes, Greece, October 4-7, 2018, Proceedings, Part III.
pp 270--279, \mn@doi{10.1007/978-3-030-01424-7_27}

\bibitem[\protect\citeauthoryear{{Tanoglidis} et~al.,}{{Tanoglidis}
  et~al.}{2022}]{Tanoglidis22}
{Tanoglidis} D.,  et~al., 2022, \mn@doi [Astronomy and Computing]
  {10.1016/j.ascom.2022.100580}, \href
  {https://ui.adsabs.harvard.edu/abs/2022A&C....3900580T} {39, 100580}

\bibitem[\protect\citeauthoryear{Wu, Kirillov, Massa, Lo  \& Girshick}{Wu
  et~al.}{2019}]{wu2019Detectron2}
Wu Y.,  Kirillov A.,  Massa F.,  Lo W.-Y.,   Girshick R.,  2019, Detectron2,
  \url{https://github.com/facebookresearch/detectron2}

\bibitem[\protect\citeauthoryear{Xie, Girshick, Doll{\'a}r, Tu  \& He}{Xie
  et~al.}{2017}]{Xie16}
Xie S.,  Girshick R.,  Doll{\'a}r P.,  Tu Z.,   He K.,  2017, in Proceedings of
  the IEEE conference on computer vision and pattern recognition. pp 1492--1500

\bibitem[\protect\citeauthoryear{Zhou et~al.,}{Zhou et~al.}{2021}]{Zhou20}
Zhou S.~K.,  et~al., 2021, Proceedings of the IEEE, 109, 820

\makeatother
\end{thebibliography}

%%%%%%%%%%%%%%%%%%%%%%%%%%%%%%%%%%%%%%%%%%%%%%%%%%

%%%%%%%%%%%%%%%%%%%%%%%%%%%%%%%%%%%%%%%%%%%%%%%%%%

%\newpage

\appendix
\section{DECam results}
\label{app:decam}

For a baseline comparison of network performances, we utilize the PhoSim dataset created and used by \cite{burke_deblending_2019}. We refer to the earlier work for a full description, but provide a brief summary here.  Crowded fields as taken with DECam are produced using the Photon Simulator code \citep{Peterson2015}. Simulations account for equipment optics \citep{Cheng2017}, telescope options \citep{DECam} and atmospheric conditions.  Spiral, elliptical and irregular galaxies are produced by sampling three-dimensional sersic profiles with additional parameters for extra morphological features. Stars are modeled as point sources and created following an initial mass function from \citep{Kroupa2001}. For both stars and galaxies, SEDs and metallicities are assigned based on physical models.  Cosmic star formation history \citep{Madau2014} is used to assign galaxy number density and population, while the distribution of stars is based on galactic latitude.  To simulated crowded fields, the galactic overdensity is boosted by a factor of 4.  A 512x512 pixel$^2$ image is produced with g,r, and z DECam bands.  Integration time and magnitude ranges are assigned to roughly correspond to DECaLs DR7 coadds \citep{Dey2019}.  In order to assign object masks, a g-band image without background is produced for every object in the field.  The PSF is configured to $\sim$1 arcsec. In total, 1000 images are produced for our training set, while an additional 250 are used for validation and another 50 as our test set for evaluation.  Each image contains roughly 150 objects.  

Here we present the results of two runs on the simulated DECam data, using the R101fpn and MViTv2 backbones.  These backbones are chosen to compare the performance of convolutional versus transformer-based architectures.  We use the same contrast scalings that were applied to the HSC data, but change the stretch parameter to 100 and Q to 10 for the Lupton and Lupton high-contrast scalings. The dynamic range of the simulated data is different from the HSC data, so the adjustment is done to make galaxy features more distinguishable.  AP scores for each configuration are shown in Table \ref{tab:decam_AP}.  We adapt the ranges for Small, Medium, and Large bounding box sizes to match those used in \cite{burke_deblending_2019}.  Overall, we find that a Lupton scaling with a ResNet backbone works best for this dataset, giving the highest AP scores for almost all categories.  This is in contrast to the results on HSC data, however we note that a transformer backbone is again more robust to contrast scalings.  Although \cite{burke_deblending_2019} use a z-scale with a R101fpn backbone, our results are different as we use a slightly altered z-scale formula in that we rescale each band by a constant $\sigma_{I}$ rather than a per-band scale factor.  This alteration makes galaxy classification performance worse (AP=29.80 compared to AP=49.6) but star classification performance better (AP=54.32 compared to AP=48.6).  The large drop in galaxy AP suggests that the R101fpn backbone is very sensitive to the contrast scaling. All other configurations result in better galaxy and star AP scores than the \cite{burke_deblending_2019} results.  Our AP scores for Small objects are lower, but Medium and Large are much higher.  For size categories, we use the same size definitions as in \cite{burke_deblending_2019}, but compute an average AP for all IOU thresholds, rather than the AP at only the lowest threshold IOU=0.5.  Thus, our results can be thought of as a kind of lower bound, as AP score tends to increase with a lower IOU threshold.
\begin{table}
\centering
\begin{tabular}{cccc}
 &  & R101fpn & MViTv2 \\
\hline
\multirow[t]{3}{*}{Galaxies} & Lupton & 65.8 & 62.5 \\
 & LuptonHC & 58.3 & 62.2 \\
 & zscale & 29.8 & 60.4 \\
 \hline
\multirow[t]{3}{*}{Stars} & Lupton & 70.1 & 68.0 \\
 & LuptonHC & 64.3 & 68.6 \\
 & zscale & 54.3 & 66.4 \\
 \hline
\multirow[t]{3}{*}{Small} & Lupton & 68.3 & 65.7 \\
 & LuptonHC & 61.8 & 65.7 \\
 & zscale & 42.3 & 63.8 \\
 \hline
\multirow[t]{3}{*}{Medium} & Lupton & 36.1 & 31.6 \\
 & LuptonHC & 29.0 & 46.7 \\
 & zscale & 16.3 & 31.7 \\
 \hline
\multirow[t]{3}{*}{Large} & Lupton & 72.6 & 54.9 \\
 & LuptonHC & 49.1 & 65.2 \\
 & zscale & 38.0 & 68.0 \\
\end{tabular}
\caption{AP scores for DECam runs. Galaxy and star AP scores improve over the results of \protect\cite{burke_deblending_2019} when different contrast scalings and backbones are applied.  Transformer-based models are more robust to contrast scalings, consistent with results on real HSC data.  }
\label{tab:decam_AP}
\end{table}

% Don't change these lines
\bsp	% typesetting comment
\label{lastpage}
\end{document}